\documentclass{article}
\usepackage{arxiv}
\usepackage{amssymb}
\usepackage{amsmath}
\usepackage{amscd}
\usepackage{graphicx}
\usepackage{graphics}
\usepackage{xcolor,graphicx}
\usepackage{xcolor}
\usepackage{csquotes}
\usepackage{mathtools}
\usepackage{caption}
\usepackage{float}
\usepackage{subcaption}
\usepackage{optidef}
 \usepackage{algorithmic}
 \usepackage{algorithm}
\usepackage{hyperref}
\usepackage[toc,page,title,titletoc]{appendix}
\usepackage{color}
\usepackage{indentfirst}
\usepackage{booktabs}
\hypersetup{colorlinks,linkcolor={blue},citecolor={blue},urlcolor={blue}}
\newcommand{\Nm}[0]{n}                       
\newcommand{\Nr}[0]{m} 
\newcommand{\Ns}[0]{p} 
\newcommand{\Np}[0]{q} 
\newcommand{\Sketch}[0]{S} 
\newcommand{\Source}[0]{B} 
\newcommand{\Orthogonal}[0]{H}
\newcommand{\Random}[0]{R}
\begin{document}
%
\title{Randomized Source Sketching for Full Waveform Inversion}
%
%
%

\author{\href{https://orcid.org/0000-0002-2080-9697}{\includegraphics[scale=0.06]{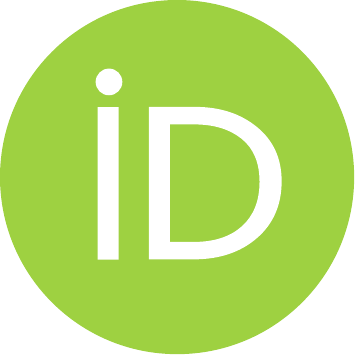}\hspace{1mm}Kamal Aghazade} \\
  Institute of Geophysics, University of Tehran, Tehran, Iran. 
  \texttt{aghazade.kamal@ut.ac.ir}
\And
  \href{http://orcid.org/0000-0003-1805-1132}{\includegraphics[scale=0.06]{orcid.pdf}\hspace{1mm}Hossein S. Aghamiry} \\
  University Cote d'Azur - CNRS - IRD - OCA, Geoazur, Valbonne, France. 
  \texttt{aghamiry@geoazur.unice.fr} 
  \And
  \href{https://orcid.org/0000-0002-9879-2944}{\includegraphics[scale=0.06]{orcid.pdf}\hspace{1mm}Ali Gholami} \\
  Institute of Geophysics, University of Tehran, Tehran, Iran.
  \texttt{agholami@ut.ac.ir} \\ 
  \And
\href{http://orcid.org/0000-0002-4981-4967}{\includegraphics[scale=0.06]{orcid.pdf}\hspace{1mm}St\'ephane Operto} \\ 
  University Cote d'Azur - CNRS - IRD - OCA, Geoazur, Valbonne, France. 
  \texttt{operto@geoazur.unice.fr}
  }
  
\renewcommand{\shorttitle}{Randomized Source Sketching FWI, Aghazade et al.}
\maketitle
\begin{abstract}
Partial differential equation (PDE) constrained optimization problems such as seismic full waveform inversion (FWI) frequently arise in the geoscience and related fields. For such problems, many observations are usually gathered by multiple sources, which form the right-hand-sides of the PDE constraint. Solving the inverse problem with such massive data sets is computationally demanding, in particular when dealing with large number of model parameters.
This paper proposes a novel randomized source sketching method for the efficient resolution of multisource PDE constrained optimization problems.
We first formulate the different source-encoding strategies used in seismic imaging into a unified framework based on a randomized sketching. To this end, the source dimension of the problem is projected in a smaller domain by a suitably defined projection matrix that gathers the physical sources in super-sources through a weighted summation. This reduction in the number of physical sources decreases significantly the number of PDE solves while suitable sparsity-promoting regularization can efficiently mitigate the footprint of the cross-talk noise to maintain the convergence speed of the algorithm. 
We implement the randomized sketching method in an extended search-space formulation of frequency-domain FWI, which relies on the alternating-direction method of multipliers (ADMM). Numerical examples carried out with a series of well-documented 2D benchmarks demonstrate that the randomized sketching algorithm reduces the cost of large-scale problems by at least one order of magnitude compared to the original deterministic algorithm.
\end{abstract}

\graphicspath{{./Figures/}}
\section{Introduction}
Partial differential equation (PDE) constrained optimization problems frequently arise in the geoscience and related fields whenever the relation between the parameters of interest and the observations (forward problem) is described by a PDE such as the wave-equation in seismology \cite{pratt1998gauss},  Poisson-type equations in DC resistivity \cite{ha2006efficient}, and Maxwell's equations in ground-penetrating radar (GPR) \cite{lavoue2014two}.
 
This paper deals with full waveform inversion (FWI), which is governed by the wave equation. 
FWI is an attractive inversion scheme with resolution capabilities at the wavelength scale to build different subsurface parameters from sparse measurements of the propagated wavefields \cite{tarantola1984inversion,pratt1998gauss,sirgue2004efficient,virieux2009overview}. 
The inverse problem is usually formulated as the minimization of a misfit distance between calculated and observed data subject to a set of equality constraints \cite{haber2000optimization}.

Traditional FWI methods (also known as reduced approaches) strictly enforce the wave-equation constraint to be satisfied during iterations by eliminating the wavefield, leading to a mono-variate unconstrained optimization problem for model parameters only. 
It is usually solved iteratively using gradient-based numerical methods in the time or frequency domain \cite{virieux2009overview}. This study is focused on the frequency-domain formulation of FWI, where seismic modelling requires the solution of large and sparse indefinite linear systems with multiple right-hand sides.
These local optimizations methods, however, are prone to cycle skipping, which drives the inversion toward spurious local minima when the initial model does not allow one to match the recorded data with an error smaller than half a period  \cite{virieux2017introduction}.

%
%

To manage the cycle-skipping issue, a new generation of algorithms known as the extended waveform inversion append extra degrees of freedom to the problem, which bring flexibility for fitting the data with inaccurate velocity models and hence reduce the cycle skipping pathology  \cite{van1997contrast,haber2000optimization,van2013mitigating,aghamiry2019improving}. 
In such extended-space approaches, wavefields or sources  are processed as optimization variables in addition to the model parameters, which implies that the PDE is solved approximately (i.e., with a relaxation) unlike the case of the reduced approach.
Among them, the wavefield reconstruction inversion (WRI) \cite{van2013mitigating} and iteratively refined-WRI (IR-WRI) \cite{aghamiry2019improving} showed promising results for accurate parameter estimation from rough starting models. WRI method is based on a penalty formulation of the constrained problem, which requires a fine-tuning of the penalty parameter, and has a slow convergence rate. 
IR-WRI overcomes these drawbacks by using the augmented Lagrangian (AL) function, which has several advantages over the penalty counterpart \cite{Bertsekas_1976_OPM}. This study  is thus focused on increasing the computational efficiency of IR-WRI based on randomized sketching methods, usually referred to as simultaneous sources or source encoding strategies.

Generally, waveform inversion is computationally intensive, which makes 3D field-scale case studies challenging. The main computational burden results from  the solution of the PDE with multiple right-hand sides. In the frequency domain, this PDE is typically solved with direct or iterative methods \cite{Duff_2017_DSS,Saad_2003_IMS}. In the latter case which may be more suitable for very large problems(more than 100 millions degrees of freedom), the computational complexity theoretically scales to the number of right-hand sides (source terms). Even if block iterative methods and recycling can speedup the processing of multiple right-hand sides \cite{Parks_2006_RKS,Jolivet_2016_BIM}, it remains beneficial to mitigate the computational burden of multisource simulations for dense acquisitions. 
Therefore, along with the developments of efficient solvers, some cost-effective strategies based on dimensionality reduction methods have been designed to handle computational challenges of waveform inversion both in terms of memory requirement and speed up purposes.  
Data decimation (\cite{sirgue2004efficient, li2012fast}), data compression, and deterministic/random stacking methods \cite{habashy2011source} are preliminary works toward this goal. Decimation methods may be unfavorable in areas with inadequate illumination \cite{castellanos2015fast}. Deterministic stacking of sources, as well as data compression methods, suffer from cross-talk noise artifacts due to the unbalanced summation of the individual sources \cite{zhang2018hybrid}. 
As a deterministic encoding strategy, the plane-wave encoding reduces the dimensionality of data by inverting the dominant components of the data after transformation into the plane-wave domain (linear Radon transform) \cite{Vigh_2008_PPF,whitmore1995imaging,zhang2005delayed,tao2013frequency,kim2020seismic}. 
However, the linear Radon transform is not able to optimally focus the seismic events in the transform domain.
This makes the performance of the plane-wave encoding highly dependent on the number of ray parameters within an appropriate range \cite{liu2006toward,tao2013frequency}.

Random projection methods consider a linear combination of the sources by applying some random codes and are known to be more effective compared with other deterministic methods  \cite{romero2000phase,krebs2009fast,berkhout2011blended,ben2011efficient,castellanos2015fast}.

Reducing the number of physical sources to a few super-sources decreases the computational cost dramatically \cite{krebs2009fast,ben2011efficient,Schiemenz_2013_AFW,castellanos2015fast},  but still two main challenges remain, namely the cross-talk noise induced by the source blending and the slow convergence rate of the algorithm when significant reduction of the source dimensionality is performed. In what regards the first issue, one may investigate different random codes \cite{schuster2011theory,ben2011efficient,haber2012effective}, re-generate random codes at each iteration \cite{haber2012effective}, use random source positions \cite{boonyasiriwat20103d}, or  employ a regularization method \cite{xue2016seismic,chen2017preserving,guitton2012attenuating}. 
Regarding the convergence issue, one may employ second-order optimization algorithms to accelerate the convergence rate \cite{anagaw2012full,castellanos2015fast,rao2017seismic}. 
 \cite{castellanos2015fast} shows that for non-convex problems, random encoding may explore areas in the model space that is not accessible to the deterministic counterpart. 
 \cite{ben2011efficient} shows that implementation of multi-frequency group inversion and partial assembling of the sources make the inversion more robust against both cross-talk and physical noises.
Finally, \cite{anagaw2014comparison} concluded that inverting partially overlapped frequency groups yields high-quality results of the source encoding method. 

The driving motivation of this paper is to increase the computational efficiency of the recent IR-WRI algorithm with randomized sketching strategies \cite{Pilanci_2015_RSC,Gower_2017_RQN}.
Randomized IR-WRI firs requires to define the associated augmented Lagrangian function in an appropriate transform domain of lower dimension. For this, the source dimension of the problem is projected by a sketching matrix in the transformed domain, whereby the physical sources are replaced by super-sources, each can be a combination of the physical sources.
For a satisfactory recovery performance of the lower-dimensional problem, the Gram matrix's expectation is associated with the sketching matrix approach identity.  In this case, the projected augmented Lagrangian function converges to the original function. 
Two categories of sketching matrices are proposed. The first category includes unstructured random matrices with independent and identically distributed (i.i.d.) entries, including those from the Gaussian and Bernoulli distributions. 
The second category includes structured random matrices that are built by randomizing orthogonal matrices.
As special cases, we use matrices associated with the Fourier transform, discrete cosine transform (DCT), discrete wavelet transform (DWT), Noiselet transform \cite{Coifman_2001_Noiselets}, and Walsh-Hadamard transform \cite{Yarlagadda_2012_HMA}.

There are three key contributions of this paper:
\begin{enumerate}
\item We develop a general algorithm for randomization of extended waveform inversion in the frame of the augmented Lagrangian method. 
\item This randomized scheme admits the existing encoding methods as special cases. 
\item We demonstrate with numerical examples that the randomized algorithm provides improved performance over existing deterministic algorithms.
\end{enumerate}
\section{Preliminaries}
The basic definitions and symbols used in this paper are collected in this section and are as follows. 
Real, complex, {integers} and natural numbers are denoted by $\mathbb{R}$,  $\mathbb{C}$, {$\mathbb{Z}$} and $\mathbb{N}$, respectively. The imaginary unit is denoted by $\iota=\sqrt{-1}$.
The number of discrete model parameters, the number of receivers, and the number of sources are denoted by $\Nm$, $\Nr$, and $\Ns$, respectively. The angular frequency is denoted by $\omega=2\pi f$. The identity matrix and diagonal matrices are denoted by $\bold{I}$ and $\text{Diag}(\bullet)$, respectively, and $\nabla^2$ is the Laplace operator.
For a matrix $\bold{X}=[\bold{x}_{1},\bold{x}_2,...,\bold{x}_{n}]$, $\bold{x}_j$ denotes the $j$th column of $\bold{X}$ and
 $\bold{x}(i)$ represents the $i$-th entry of the column vector $\bold{x}$.  The conjugate transpose of a matrix/vector is denoted by superscript $T$.    
The Frobenius norm of a matrix  $\bold{X}$ is defined as  $\|\bold{X}\|_\mathrm{F} \equiv \sqrt{\sum_{i,j}|X(i,j)|^{2}}$ and for two matrices $\bold{X}$ and $\bold{Y}$ of same dimension the inner product is defined as $\langle\bold{X},\bold{Y}\rangle= \text{tr}(\bold{X}\bold{Y}^T)= \sum_{i,j}\bar{X}(i,j)Y(i,j)$ where the bar shows complex conjugate.

\section{Method}
In this section, we develop a new algorithm of IR-WRI \cite{aghamiry2019improving} based upon a randomized sketching scheme.
We formulate the augmented Lagrangian method in the transform domain of lower dimension and reviews a quite general formulation that admits existing source encoding strategies as special cases.

\subsection{Extended FWI with augmented Lagrangian method}
Multisource FWI problem with a general regularization term $\mathcal{R}(\bold{m}):\mathbb{R}^{\Nm\times 1}\to \mathbb{R}$ can be expressed as the following equality-constrained bi-convex optimization problem \cite{haber2000optimization,aghamiry2019compound}: 
\begin{mini} 
{\bold{U},\bold{m}\in \mathcal{B}}{\mathcal{R}(\bold{m})} {} \quad \addConstraint {\bold{A}(\bold{m})\bold{U}}{=\bold{\Source}}
\addConstraint {\bold{P}\bold{U}}{=\bold{D}},
{\label{main}}
\end{mini}
where the columns of $\bold{U} \in \mathbb{C}^{\Nm\times \Ns}$ gather $\Ns$ wavefields represented by vectors of dimension $\Nm$.
The columns of $\bold{D} \in \mathbb{C}^{\Nr\times \Ns}$ contain the associated $\Ns$ observation vectors of dimension $\Nr$.
The matrix $\bold{P}  \in \mathbb{R}^{\Nr\times \Nm}$ is the observation operator that samples the wavefields at the receiver locations. Through this paper, we assume a fixed-spread acquisition where $\bold{P}$ is independent of the sources. 
The matrix $\bold{A(m)} \in \mathbb{C}^{\Nm\times \Nm}$ is the discretized Helmholtz operator with appropriate boundary conditions \cite{chen2013optimal} and the columns of $\bold{\Source}  \in \mathbb{C}^{\Nm\times \Ns}$ gather the source terms. 
Finally, the vector $\bold{m} \in \mathbb{R}^{\Nm\times 1}$ gathers the model parameters (squared slowness), and $\mathcal{B}$ is a convex set defined by known lower and upper bounds of the model.
In this paper, we develop our algorithm using the frequency-domain formulation of FWI, but the resulting algorithm can also be applied in the time domain. 

The traditional FWI solves \eqref{main} by eliminating the wavefields, i.e.  $\bold{U}=\bold{A(m)}^{-1}\bold{\Source}$, leading to the following unconstrained monovariate optimization problem: 
\begin{equation} \label{pratt_FWI}
\underset{\bold{m}\in \mathcal{B}} {\text{minimize}}  ~~~~~ \mathcal{R}(\bold{m})+\frac{\mu}{2}\|\bold{P}\bold{A}(\bold{m})^{-1}\bold{\Source}-\bold{D}\|_\mathrm{F}^2,
\end{equation}
where $\mu>0$ is the penalty term.
Gradient-based optimization methods are usually used to minimize this objective function where the gradient of the misfit function is efficiently computed with the adjoint-state method \cite{Plessix_2006_RAS}. 
%
However, there is a general agreement on the sensitivity of FWI to the accuracy of the starting model when gradient-based optimization methods are used to solve it.
The AL method was proposed to mitigate this sensitivity.

The AL function of the constrained problem \eqref{main} is given by \cite{powell1969method,aghamiry2019improving}
\begin{align}\label{AL_deterministic}
\mathcal{L}(\bold{U},\bold{m},\bold{\Lambda},\bold{\Gamma}) &=  \mathcal{R}(\bold{m}) +\frac{\alpha}{2} \|\bold{A(m)U}-\bold{\Source}\|_\mathrm{F}^2 
+\frac{\beta}{2}\|\bold{PU}-\bold{D}\|_\mathrm{F}^2 - \langle\bold{\Lambda},\bold{A(m)U}-\bold{\Source}\rangle - \langle\bold{\Gamma},\bold{PU}-\bold{D}\rangle,
\end{align}
where $\alpha,\beta>0$ are penalty parameters. The matrices $\bold{\Lambda}\in \mathbb{C}^{\Nm \times \Ns}$ and $\bold{\Gamma}\in \mathbb{C}^{\Nr \times \Ns}$ gather the Lagrange multipliers associated with the PDE constraints and the observation constraints, respectively, each column of these matrices being tied to a particular source.
The AL function \eqref{AL_deterministic} recast FWI as a the following min-max problem:
\begin{equation}\label{minimax_deter}
\underset{\bold{U},\bold{m}} {\min} ~~ \underset{\bold{\Lambda},\bold{\Gamma}} {\max} ~~~ \mathcal{L}(\bold{U},\bold{m},\bold{\Lambda},\bold{\Gamma}). 
\end{equation}
IR-WRI solves \eqref{minimax_deter} iteratively with the alternating direction method of multipliers (ADMM) \cite{boyd2011distributed}, where 
each primal and dual variables are updated in alternating mode as
\begin{subequations}
\label{ADMM}
\begin{align}
& \mathbf{U}_{k+1}=\arg\min_{\mathbf{U}}~~\mathcal{L}(\bold{U},\bold{m}_k,\bold{\Lambda}_k,\bold{\Gamma}_k), \label{ADMM_u} \\ 
&\mathbf{m}_{k+1}= \arg\min_{\mathbf{m}} ~~\mathcal{L}(\bold{U}_{k+1},\bold{m},\bold{\Lambda}_k,\bold{\Gamma}_k), \label{ADMM_m}  \\ 
&\bold{\Lambda}_{k+1}=\bold{\Lambda}_{k} - \alpha(\mathbf{A}(\bold{m}_{k+1}) \mathbf{U}_{k+1} -\bold{\Source}), \label{ADMM_bk} \\
&\bold{\Gamma}_{k+1}=\bold{\Gamma}_{k} - \beta(\mathbf{P} \mathbf{U}_{k+1}-\mathbf{D}), \label{ADMM_dk}
\end{align}
\end{subequations}
beginning with $\bold{\Lambda}_0= \bold{0}$ and $\bold{\Gamma}_0= \bold{0}$, and $k$ stands for the iteration number. 
The dual updating in \eqref{ADMM_bk} and \eqref{ADMM_dk} can be interpreted as gradient ascent steps with respect to $\bold{\Lambda}$ and $\bold{\Gamma}$ (with step lengths $\alpha$ and $\beta$) applied to the dual function or to the AL function while the primal variables are kept fixed \cite{Bertsekas_1976_OPM}. 
We now provide the solution of the wavefield and model subproblems.
\subsubsection{Wavefield subproblem}
Solving the wavefield subproblem \eqref{ADMM_u} by differentiating the AL function with respect to $\bold{U}$ and equating the result to zero gives the closed-form expression of $\bold{U}_{k+1}$  as the solution of the following linear system with multiple right-hand sides
\begin{equation} \label{wsub}
(\alpha\mathbf{A}_k^{\mathrm{T}}\mathbf{A}_k + \beta\bold{P}^{\mathrm{T}}\bold{P})\bold{U}_{k+1}=\alpha\mathbf{A}_k^{\mathrm{T}}(\bold{\Source}-\frac{1}{\alpha}\bold{\Lambda}_{k}) + \beta\bold{P}^{\mathrm{T}}(\bold{D}-\frac{1}{\beta}\bold{\Gamma}_k).
\end{equation}
The computational burden of solving this system is proportional to the number of right-hand sides when iterative solvers, which are the only choice for large-scale problems, are used. 
In section \ref{mainsec} we use randomized sketching methods to reduce the dimensionality of the problem and decrease the computational burden of the inverse problem.  
\subsubsection{Model subproblem}
The objective function for the model parameters in \eqref{ADMM_m} reduces to
\begin{equation} \label{WRI_misfit}
\underset{\bold{m}\in \mathcal{B}} {\text{minimize}}~\mathcal{R}(\bold{m})+\frac{\alpha}{2} \|\bold{A(m)U}_{k+1}-\bold{\Source}+\frac{1}{\alpha}\bold{\Lambda}_{k}\|_\mathrm{F}^2.
\end{equation}
Compared to the classical FWI, eq. \ref{pratt_FWI}, in which the nonquadratic misfit function is defined in the data space, the misfit function is defined in the source space in \eqref{WRI_misfit} and is quadratic with respect to the model parameters. Considering the special structure of the Helmholtz operator
\begin{equation}
\bold{A(m)}=\nabla^2+\omega^2\text{Diag}(\bold{m}),
\end{equation}
we can rewrite \eqref{WRI_misfit} in the following equivalent form 
\begin{equation} \label{WRI_misfit2}
\underset{\bold{m}\in \mathcal{B}} {\text{minimize}}~\mathcal{R}(\bold{m})+\frac{\alpha\omega^2}{2}\sum_{j=1}^{\Ns} \|\text{Diag}(\bold{u}^{k+1}_j)\bold{m}-\bold{y}_j^k\|_2^2,
\end{equation}
where $\bold{y}_j^k=\frac{1}{\omega^2}(\bold{b}_j-\frac{1}{\alpha}\boldsymbol{\lambda}_j^{k}-\nabla^2\bold{u}^{k+1}_j)$, and  $\bold{b}_j$, $\boldsymbol{\lambda}_j^{k}$ and $\bold{u}^{k+1}_j$ are the $j$th column of matrices $\bold{\Source}$, $\bold{\Lambda}_{k}$ and $\bold{U}_{k+1}$, respectively.
The optimization problem \eqref{WRI_misfit2} can be solved efficiently with different regularization function, which may force smoothness, sparseness, blockiness. Alternatively, a combination of them can be used to compute the regularized solution \cite{Gholami_2013_BCT}. Applications of such regularizers in IR-WRI are reviewed in \cite{aghamiry2019compound,Aghamiry_2020_FWI}.
In this study, this regularization term is essential to mitigate the footprint of data over-fitting resulting from cross-talk noise.  
\section{IR-WRI with Source Sketching} \label{mainsec}
For two matrices $\bold{X},\bold{Y}\in \mathbb{C}^{\bullet \times \Ns}$ and sketching matrix $\bold{\Sketch}\in \mathbb{C}^{\Ns\times q}$ such that $\bold{\Sketch\Sketch}^{\mathrm{T}}=\bold{I}$ (the identity matrix) we have
\begin{subequations}
\begin{align} 
\langle\bold{X},\bold{Y}\rangle 
&= \langle\bold{X},\bold{Y}\bold{I}\rangle  \label{dot1}\\
&= \langle\bold{X},\bold{Y}\bold{\Sketch\Sketch}^{\mathrm{T}}\rangle \label{dot2}\\
&= \langle\bold{X}\bold{\Sketch},\bold{Y}\bold{\Sketch}\rangle \label{dot3}\\
&= \langle\bold{\hat{X}},\bold{\hat{Y}}\rangle, \label{dot4}
\end{align}
\end{subequations}
where the hat shows row-transformed matrix obtained by right multiplication by the sketching matrix $\bold{\Sketch}$.
Furthermore, the AL function in \eqref{AL_deterministic} can be written as
\begin{align} \label{AL_dot}
\mathcal{L}(\bold{U},\bold{m},\bold{\Lambda},\bold{\Gamma}) &=\mathcal{R}(\bold{m})+\frac{\alpha}{2} \langle\delta\bold{\Source},\delta\bold{\Source}\rangle + \frac{\beta}{2}\langle\delta\bold{D},\delta\bold{D}\rangle-\langle\bold{\Lambda},\delta\bold{\Source}\rangle -\langle\bold{\Gamma},\delta\bold{D}\rangle,
\end{align}
where $\delta\mathbf{D}=\mathbf{P} \mathbf{U}-\mathbf{D}$ and $\delta\bold{\Source}=\bold{A}(\bold{m}) \bold{U}-\bold{\Source}$. 
Using \eqref{dot3}, the AL function can be defined in the transform domain as
\begin{align}\label{AL_trans}
\hat{\mathcal{L}}(\hat{\bold{U}},\bold{m},\hat{\bold{\Lambda}},\hat{\bold{\Gamma}}) &=  \mathcal{R}(\bold{m}) + \frac{\alpha}{2} \|\bold{A(m)}\hat{\bold{U}} -\hat{\bold{\Source}}\|_\mathrm{F}^2 
 +\frac{\beta}{2}\|\bold{P}\hat{\bold{U}}-\hat{\bold{D}}\|_\mathrm{F}^2 -\langle\hat{\bold{\Lambda}},\bold{A(m)}\hat{\bold{U}}-\hat{\bold{\Source}}\rangle -\langle\hat{\bold{\Gamma}},\bold{P}\hat{\bold{U}}-\hat{\bold{D}}\rangle,
\end{align}
where $\hat{\bold{\Source}}={\bold{\Source}}\bold{\Sketch}$, $\hat{\bold{D}}={\bold{D}}\bold{\Sketch}$, $\hat{\bold{U}}={\bold{U}}\bold{\Sketch}$, $\hat{\bold{\Lambda}}={\bold{\Lambda}}\bold{\Sketch}$, and $\hat{\bold{\Gamma}}={\bold{\Gamma}}\bold{\Sketch}$ and the IR-WRI algorithm in \eqref{ADMM} can be readily applied to the AL function in the transform domain. 
For $\bold{\Source}\in \mathbb{C}^{\Nm\times \Ns}$ and ${\bold{\Sketch}}\in \mathbb{R}^{\Ns\times q}$, $\hat{\bold{\Source}}\in \mathbb{C}^{\Nm\times q}$ thus for $q \ll \Ns$ we expect significant computational saving if we perform optimization over $\hat{\mathcal{L}}$ rather than ${\mathcal{L}}$ while at the same time being able to maintain the good quality of the estimates.
This indeed results because, if $q\ll \Ns$, the sketched PDE \eqref{wsub} is of a much smaller dimension than the original one, and hence easier to solve. 
However, for $q< \Ns$ the equality $\bold{\Sketch\Sketch}^{\mathrm{T}}=\bold{I}$ is no longer be satisfied. Thus,
we deal with cases where $\bold{\Sketch\Sketch}^{\mathrm{T}}\approx\bold{I}$, which requires us to work with an approximate AL function $\hat{\mathcal{L}}$.
For a fixed matrix $\bold{\Sketch}$, one can  build the projected source/data matrices $\hat{\bold{\Source}}/\hat{\bold{D}}$ once and for all before the inversion and then apply IR-WRI to the projected source and data. 
In this case, however, the imprint of cross-talk noises will propagate into the model update unless $q$ is sufficiently large such that $\bold{\Sketch\Sketch}^{\mathrm{T}}\approx\bold{I}$.
To address the cross-talk noise effects while maintaining the computational efficiency of the algorithm at the desired level, we apply a randomization scheme. That is, we regenerate the projection matrix $\bold{\Sketch}$ at each IR-WRI iteration. In this case, however, we need to keep the dual parameters in the original space like the model parameters.
Thus, we rewrite the transformed AL function \eqref{AL_trans} in an equivalent form by using the equality \eqref{dot2} for the Lagrangian terms 
\begin{align}\label{AL_trans_eq}
\hat{\mathcal{L}}(\hat{\bold{U}},\bold{m},\bold{\Lambda},\bold{\Gamma}) &=  \mathcal{R}(\bold{m}) + \frac{\lambda}{2} \|\bold{A(m)}\hat{\bold{U}} -\hat{\bold{\Source}}\|_\mathrm{F}^2 +\frac{\gamma}{2}\|\bold{P}\hat{\bold{U}}-\hat{\bold{D}}\|_\mathrm{F}^2 
-\langle{\bold{\Lambda}},(\bold{A(m)}\hat{\bold{U}}-\hat{\bold{\Source}})\bold{\Sketch}^{\mathrm{T}}\rangle -\langle{\bold{\Gamma}},(\bold{P}\hat{\bold{U}}-\hat{\bold{D}})\bold{\Sketch}^{\mathrm{T}}\rangle.
\end{align}
The randomized IR-WRI is summarized in Algorithm \ref{alg1}.
Generally, the sketching matrix $\bold{\Sketch}$ and the scalar parameter $\Np$ can be seen as additional parameters of the randomized IR-WRI compared to the traditional IR-WRI.
In the following subsections, we provide possible choices of the sketching matrix $\bold{\Sketch}$, which are built either randomly or from randomized orthonormal matrices.

\begin{algorithm}
 \begin{algorithmic}[1]
 \small
 \caption{IR-WRI with Randomized Source Sketching.} \label{alg1}
 \REQUIRE $\bold{D}$, $\bold{\Source}$
 \STATE set $\bold{{\Lambda}}_0=\bold{0}$, $\bold{{\Gamma}}_0=\bold{0}$
 \vspace{0.1cm}
 \REPEAT
 \STATE Sample an independent copy~ $\bold{\Sketch}$
 \vspace{0.1cm}
 \STATE Set $\mathbf{\hat{\Source}}=\bold{\Source\Sketch}, \bold{\hat{D}}=\bold{D\Sketch},\bold{\hat{\Lambda}}_{k}=\bold{{\Lambda}}_{k}\bold{\Sketch}, \bold{\hat{\Gamma}}_{k}=\bold{{\Gamma}}_{k}\bold{\Sketch}$
 \vspace{0.1cm}
 \STATE $\mathbf{\hat{U}}_{k+1}=\bold{H}^{-1}[\alpha\mathbf{A}_k^{\mathrm{T}}(\mathbf{\hat{\Source}}-\frac{1}{\alpha}\bold{\hat{\Lambda}}_{k}) + \beta\bold{P}^{\mathrm{T}}(\bold{\hat{D}}-\frac{1}{\beta}\bold{\hat{\Gamma}}_k)]$
 \vspace{0.1cm}
 \STATE $\mathbf{m}_{k+1}= \text{argmin}_{\mathbf{m}}~ \mathcal{R}(\bold{m})+\frac{\alpha}{2} \|\bold{A(m)\hat{U}}_{k+1}-\bold{\hat{\Source}}+\frac{1}{\alpha}\bold{\hat{\Lambda}}_{k}\|_\mathrm{F}^2$
 \vspace{0.1cm}
\STATE $\bold{{\Lambda}}_{k+1}=\bold{{\Lambda}}_{k} - \alpha(\mathbf{A}(\bold{m}_{k+1}) \mathbf{\hat{U}}_{k+1} - \mathbf{\hat{\Source}})\bold{\Sketch}^{\mathrm{T}}$
\vspace{0.1cm}
\STATE $\bold{{\Gamma}}_{k+1}=\bold{{\Gamma}}_{k} - \beta(\mathbf{P} \mathbf{\hat{U}}_{k+1} - \mathbf{\hat{D}})\bold{\Sketch}^{\mathrm{T}}$
\vspace{0.1cm}
\STATE $k = k+1$.
\UNTIL{stopping conditions are satisfied.}
\end{algorithmic}
\end{algorithm}
\subsection{Random sketching matrices} 
We would like the random matrix $\bold{\Sketch}$ to satisfy 
\begin{equation} \label{EX}
\mathbb{E}(\bold{\Sketch}\bold{\Sketch}^{\mathrm{T}})=\bold{I},
\end{equation}
where $\mathbb{E}$ denotes expectation over random matrix $\bold{\Sketch}$ with a pre-defined distribution.
A truncated expectation has been widely used for stochastic estimation of the trace of a matrix  \cite{hutchinson1989stochastic}.
The stochastic trace estimation has also been applied to randomized/stochastic FWI problems \cite{van2011seismic,louboutin2021ultra}. 
In the following, we review different forms of $\bold{\Sketch}$, which are referred to as random projection matrices because each element of them is randomly generated.

\subsubsection{Gaussian Sketch}
The $\Ns\times \Np$ matrix $\bold{\Sketch}$ reads 
\begin{equation}
\bold{\Sketch}(i,j)= \frac{1}{\sqrt{q}} G_{ij},
\end{equation}
where $G_{ij}$ is identically and independently sampled from a standard normal distribution $\mathcal{N}(0, 1)$.

\subsubsection{Bernoulli/Rademacher Sketch}
The $\Ns\times \Np$ Bernoulli sketching matrix $\bold{\Sketch}$ is 
\begin{equation}
\bold{\Sketch}(i,j)= \frac{1}{\sqrt{q}}B_{ij},
\end{equation}
where $B_{ij}$ is identically and independently sampled from a symmetric Bernoulli distribution $\mathrm{Pr}(B_{ij}=\pm 1)=\frac{1}{2}$.

\subsubsection{Random phase Sketch}
In this case the $\Ns \times \Np$ sketching matrix $\bold{\Sketch}$ is defined as 
\begin{equation}
\bold{\Sketch}(i,j)= \frac{1}{\sqrt{q}} e^{\iota \varphi_{ij}},
\end{equation}
where $\varphi_{ij}$ are random numbers in the interval $[0,2\pi]$.

\subsubsection{Count Sketch}
In this case the $\Ns\times \Np$ matrix has a single randomly chosen non-zero entry in each row \cite{Charikar_2004_FFI}.
Each non-zero entry is a Rademacher random variable $\pm 1$ with probability 1/2. 
Intuitively, the product $\bold{\Source}\bold{\Sketch}$ for count sketch matrix $\bold{\Sketch}$ can be computed in three steps \cite{Wang_2015_APG}: First, hash each source with a discrete value uniformly sampled from $\{1,2,\cdots,\Np\}$. Second, multiply each source by a Rademacher random variable $\pm 1$ with probability 1/2. Third, sum up the sources with the same hash value.
This is similar to a random sampling of the sources as in section \ref{diarc}. However, the advantage of count sketching is that all of the sources are used at each iteration, but unlike the Gaussian, Bernoulli, and phase random projections, each blended source is a combination of a subset of the physical sources.

\subsection{Randomized orthogonal matrices} 
In order to define a randomized {orthogonal} matrix, we begin by an {orthogonal}  matrix $\bold{\Orthogonal}$ of size $\Ns\times \Ns$ then sketching matrix $\bold{\Sketch}$ is defined as
\begin{equation} \label{Sketch_mat}
\bold{\Sketch} = \bold{\Pi}\bold{\Orthogonal}\bold{\Random},
\end{equation}
where each column of the $\Ns\times \Np$ matrix $\bold{\Random}$ is chosen 
{randomly} from the set of all $\Ns$ canonical basis vectors (the columns of the identity matrix). 
The matrix $\bold{\Pi}$ can be a permutation matrix that reorders the rows of the matrix $\bold{\Orthogonal}$ randomly,  or it can be a diagonal matrix with diagonal entries equal to +1 or -1 with probability 1/2 \cite{Pilanci_2015_RSC}.
In what follows, we review some orthogonal matrices that are commonly used in applications.

\subsubsection{Identity matrix} \label{diarc}
The simplest choice for $\Orthogonal$ is the identity matrix, $\bold{\Orthogonal=I}$. In this case, in each iteration of the randomized IR-WRI we use a number of $\Np$ sources that are selected uniformly at random from the $\Ns$ physical sources. 
\subsubsection{Discrete Fourier transform}
The matrix $\bold{\Orthogonal}$ reads 
\begin{equation}
\bold{\Orthogonal}(i,j) = \frac{1}{\sqrt{q}} e^{\frac{\iota 2\pi (i-1)(j-1)}{\Ns}},~~i,j=1,\cdots,\Ns.
\end{equation}
\subsubsection{Discrete cosine transform}
The matrix $\bold{\Orthogonal}$ reads 
\begin{equation}
\bold{\Orthogonal}(i,j) = \left\{
\begin{array}{ll}
\frac{1}{\sqrt{\Ns}}\cos(\frac{\pi (2i-1)(j-1)}{2\Ns})&\text{if}~ j=1,\\
\sqrt{\frac{2}{\Ns}}\cos(\frac{\pi (2i-1)(j-1)}{2\Ns})&\text{if}~ j\neq 1,
\end{array} \right.
\end{equation}
$i,j=1,\cdots,\Ns$.
The discrete cosine transform (DCT) is a widely used tool for signal compression. 
\begin{figure}[!t]
\begin{center}
\includegraphics[scale=0.9,trim={0.5cm 0 0 1cm},clip]{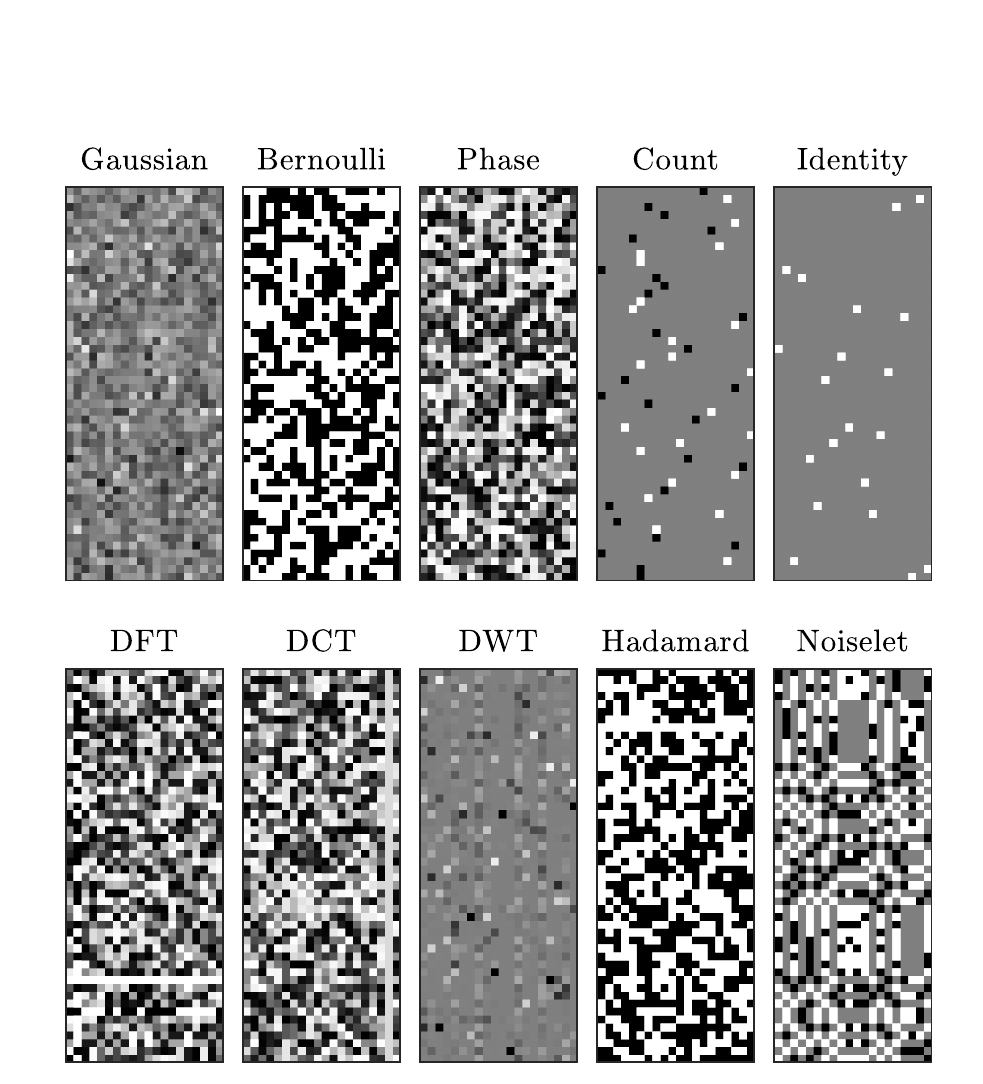}
\end{center}
\caption{Samples of sketching matrices for $p=50$ and $q=20$. The color range between -1 (black) and +1 (white).} \label{S_matrices}
\end{figure}
\subsubsection{Discrete wavelet transform (DWT)}
 It is another effective transform for signal compression and multiscale analysis \cite{Mallat_1999_AWT}. 
The basis functions for DWT associated with the quadrature mirror filters $h(k)$ and $g(k)$  are constructed at the scale $j\in \mathbb{Z}$ and location $k\in \mathbb{Z}$ as $\varphi_{j_0,k}(x)=\sqrt{2^{j_0}}\varphi(2^{j_0}x-k)$ and $\psi_{j_0,k}(x)=\sqrt{2^{j}}\psi(2^{j}x-k)$ where the scaling function $\varphi(x)$ and the wavelet function $\psi(x)$ are obtained by solving the following two-scale difference equations
\begin{equation}
\left\{
\begin{array}{ll}
\varphi(x)=\sqrt{2}\sum_k h(k)\varphi(2x-k),\\
\psi(x)=\sqrt{2}\sum_k g(k)\varphi(2x-k).
\end{array} \right.
\end{equation}
   
The DWT matrix requires $p$ to be a power of 2. For $p$ that is not a power of 2, one may take a random subsample of rows and columns of the matrix that is a power of 2.
\subsubsection{Noiselet transform}
It is originally presented in \cite{Coifman_2001_Noiselets} as a measuring matrix for compressed sensing of signals that have a sparse representation in the Haar wavelet domain because Noiselets are maximally incoherent to the Haar basis and have a fast implementation algorithm. 
The Noiselet matrix of order $p$ is defined recursively by
\begin{equation}
\bold{\Orthogonal}_p = 
\frac{1-\iota}{2}\bold{\Pi}
\begin{bmatrix}
\iota\bold{\Orthogonal}_{p/2} & \bold{\Orthogonal}_{p/2} \\
\bold{\Orthogonal}_{p/2} & \iota\bold{\Orthogonal}_{p/2}
\end{bmatrix},
\end{equation}
beginning with $\bold{\Orthogonal}_1=\bold{1}$, where $\bold{\Pi}$ is a permutation matrix such that for $\bold{y=\Pi x}$ we have $\bold{y}(2i-1)=\bold{x}(i)$ and $\bold{y}(2i)=\bold{x}(i+\frac{n}{2})$ for $i=1,2,...,\frac{p}{2}$.
For example, the matrix $\bold{\Pi}$ for $p=8$ can be explicitly expressed as
\begin{equation}
\bold{\Pi}=
\begin{bmatrix}
1 & 0 & 0 & 0 & 0 & 0 & 0 & 0 \\
0 & 0 & 0 & 0 & 1 & 0 & 0 & 0 \\
0 & 1 & 0 & 0 & 0 & 0 & 0 & 0 \\
0 & 0 & 0 & 0 & 0 & 1 & 0 & 0 \\
0 & 0 & 1 & 0 & 0 & 0 & 0 & 0 \\
0 & 0 & 0 & 0 & 0 & 0 & 1 & 0 \\
0 & 0 & 0 & 1 & 0 & 0 & 0 & 0 \\
0 & 0 & 0 & 0 & 0 & 0 & 0 & 1
\end{bmatrix}.
\end{equation}
The Noiselet thus requires $p$ to be a power of 2. For $p$ that is not a power of 2, one may take a random subsample of rows and columns of the matrix that is a power of 2.

\begin{figure}[!t]
\begin{center}
\includegraphics[scale=0.9]{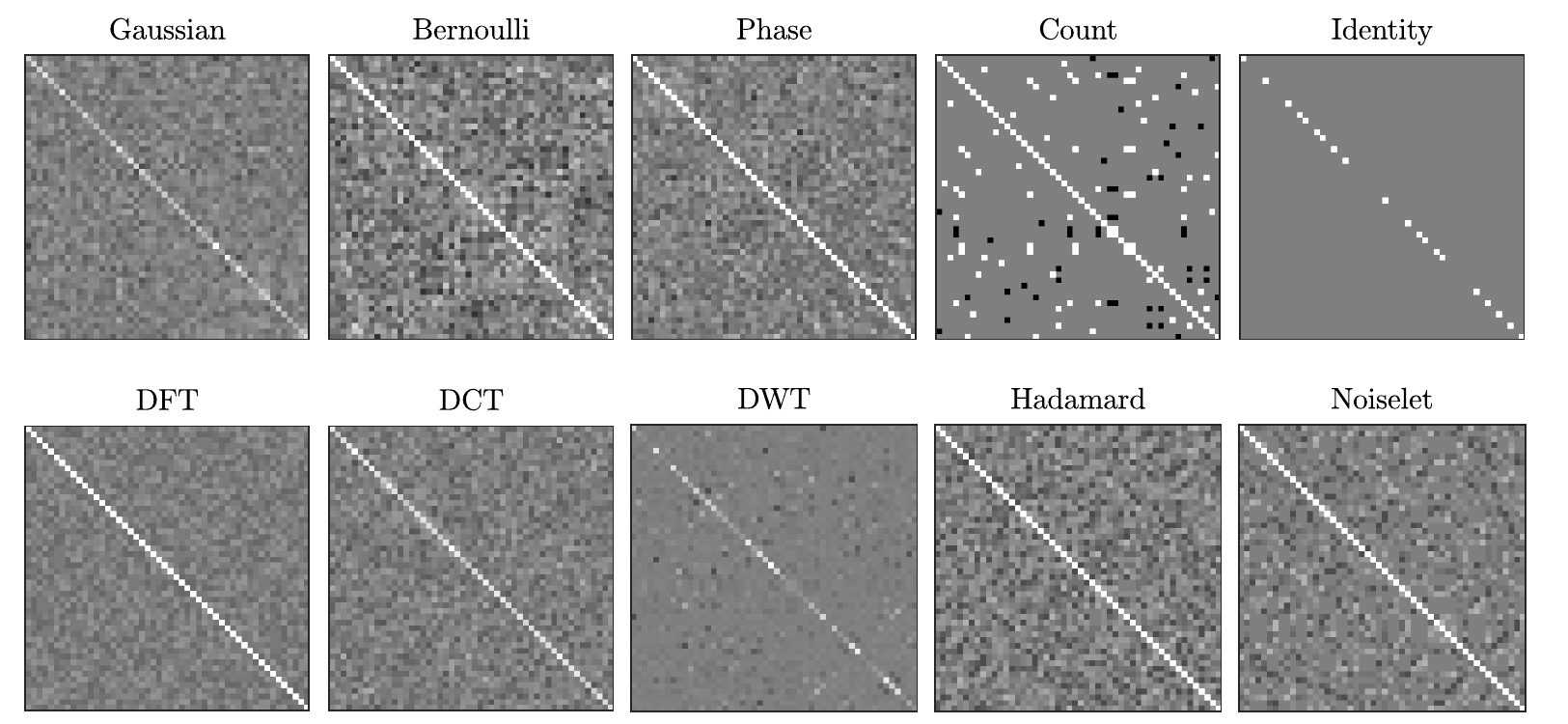}
\end{center}
\caption{Gram matrices associated with the samples of sketching matrices shown in Fig.~\ref{S_matrices} ($p=50$ and $q=20$). The color scale ranges between -1 (black) and +1 (white).} \label{Gram_matrices}
\end{figure}
\subsubsection{Hadamard transform}
The Walsh-Hadamard matrix of order $p$ is defined recursively by \cite{Yarlagadda_2012_HMA} as
\begin{equation}
\bold{\Orthogonal}_p = 
\frac{1}{\sqrt{2}}
\begin{bmatrix}
\bold{\Orthogonal}_{p/2} & \bold{\Orthogonal}_{p/2} \\
\bold{\Orthogonal}_{p/2} & -\bold{\Orthogonal}_{p/2}
\end{bmatrix},
\end{equation}
beginning with $\bold{\Orthogonal}_1=1$. This matrix also requires $p$ to be a power of 2. For $p$ that is not a power of 2, one may take a random subsample of rows and columns of the matrix that is a power of 2.
Similar to the Bernoulli and count sketch matrices, the entries of the Walsh-Hadamard matrix are either -1 or +1. However, the rows and columns of the Walsh-Hadamard matrix are mutually orthogonal.

\begin{figure}[!t]
\begin{center}
\includegraphics[scale=0.9]{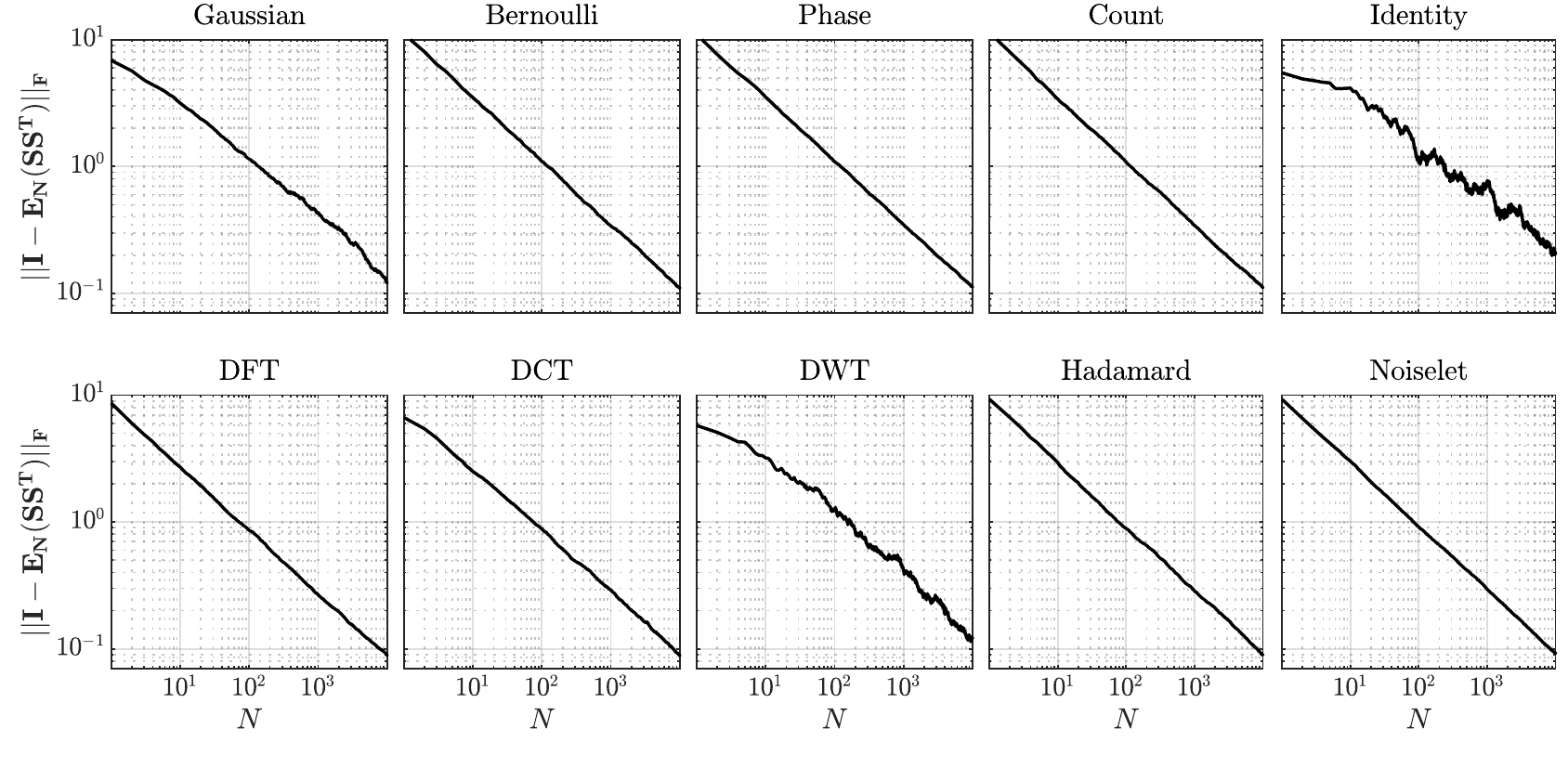}
\end{center}
\caption{The error between identity matrix and truncated expectation of sketching matrices for $p=50$ and $q=20$ versus level of truncation $N${(number of realizations)}.} \label{E_curves}
\end{figure}

 
Figs. \ref{S_matrices} and \ref{Gram_matrices} compare samples of sketching matrices $\bold{\Sketch}$ and associated Gram matrices $\bold{\Sketch}\bold{\Sketch}^T$ for the aforementioned methods.
Since we require the expectation of the Gram matrix to approach the identity matrix, the associated error curve for different sketching matrices is plotted in Fig. \ref{E_curves}. Except for the Identity and DWT, other matrices are well behaved with a little difference in their convergence rate.
\section{Numerical Examples}
We assess the performance of the  IR-WRI method with randomized source sketching for three 2D mono-parameter benchmark models: SEG/EAGE Overthrust \cite{Aminzadeh_1997_DSO}, 2004 BP model \cite{Billette_2004_BPB}, and 2007 BP model created by Hemang Shah from BP Exploration Operation Company in 2007.  Frequency-domain finite-difference modeling with a 9-point stencil and perfectly-matched layer (PML) \cite{chen2013optimal} is used for discretising the wave-equation in this paper. Also, a fixed penalty parameter is selected for all the numerical tests following the instruction given in  \cite{aghamiry2019improving}. Additionally, we utilized bound constraints introduced in \cite{aghamiry2019implementing,aghamiry2019compound} with the minimum and maximum of the true model as lower and upper bounds. Furthermore, in order to make the performance of different (deterministic and stochastic) methods comparable, we normalize the sketching matrix $\bold{\Sketch}$ at each iteration such that its maximum singular value is unity. 

We also utilize the traditional plane-wave encoding. It has been applied to both FWI, and WRI in a deterministic setting \cite{Vigh_2008_PPF,zhang2005delayed,tao2013frequency,kim2020seismic}. It is based on transforming the source dimension by the linear Radon transform and then inverting the coefficients associated with a predefined range of angles. The plane wave encoding can be properly formulated with a fixed sketching matrix (see Appendix \ref{PW_sec}). 
%
%
However, we found that, for different models, satisfactory results can be obtained only when the coefficients span over a wide range of angles, which translates to a large number of plane-wave sources. In order to increase the computational efficiency while maintaining the reconstruction accuracy, we use a randomized plane-wave encoding; in which only $q\ll p$ random plane-wave sources are used at each iteration.
%
%
%
%
\begin{figure}[!t]
\begin{center}
\includegraphics[scale=0.65]{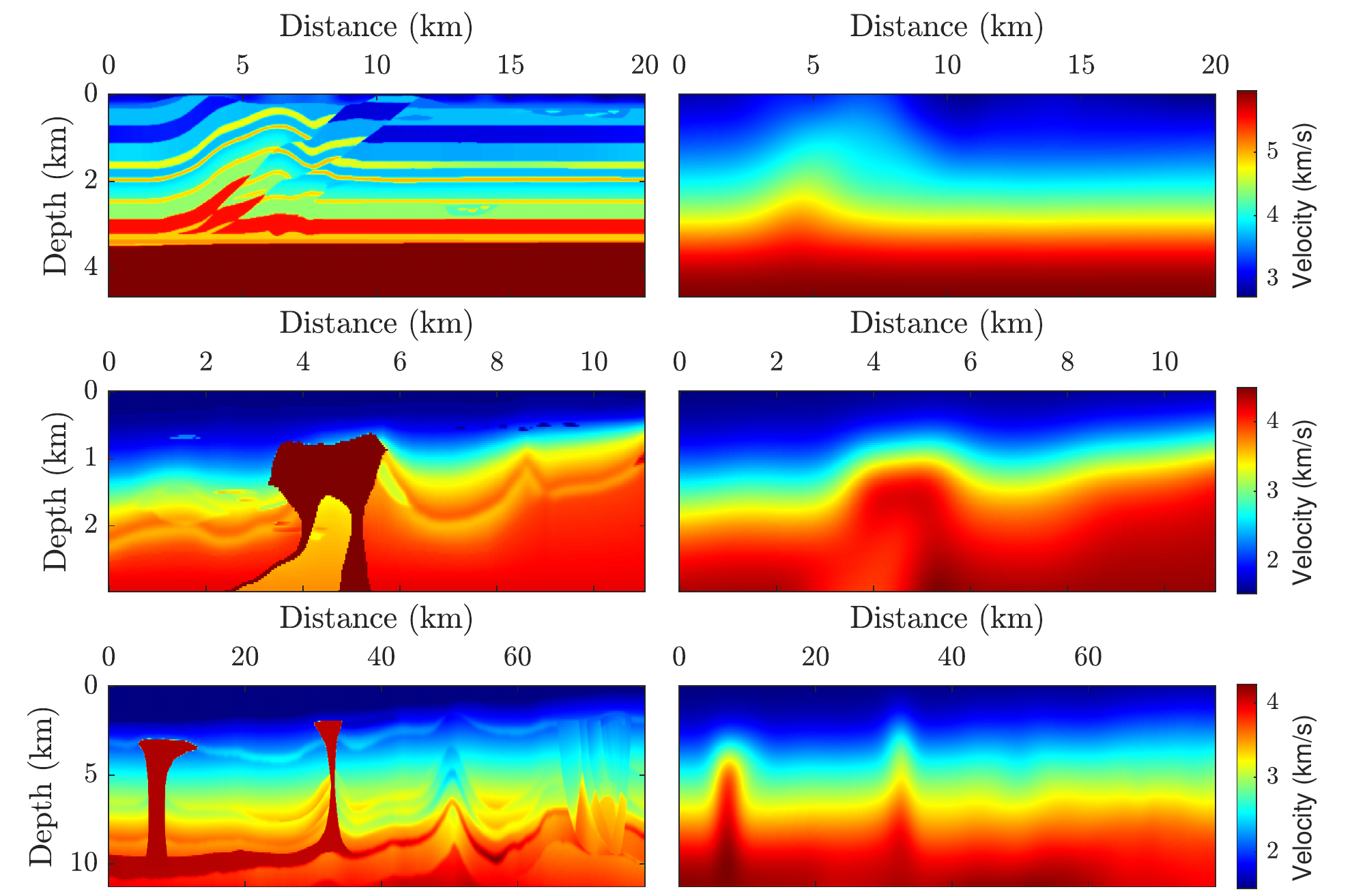}
\end{center}
\caption{The SEG/EAGE Overthrust velocity model (top), 2004 BP velocity model (middle), and 2007 BP velocity model (bottom). The left column shows the true model, and the right column shows the initial models used for the inversion.} \label{velmodels}
\end{figure}

\subsection{SEG/EAGE Overthrust exmaple}
We assess the performance of IR-WRI with randomized source sketching against a 2D section of the 3D SEG/EAGE Overthrust model (Fig. \ref{velmodels}, top left). The model includes 801 $\times$ 187 grid points with a grid interval of 25~m in both spatial directions. The surface acquisition is conducted with $p=134$ sources and $m=801$ receivers with an interval of 150~m and 25~m, respectively. A smooth version of the true velocity model is used as the initial model (Fig. \ref{velmodels}, top right), and the inversion is performed over two frequency paths, [3-6.5]~Hz and [3-13]~Hz with a frequency interval of 0.5~Hz in which the final estimated model of the first path is used as an initial model for the second path. Moreover, the maximum number of iterations per frequency is set to 10, except for the first two frequencies for which we perform 30 iterations. We perform the inversion with both deterministic and randomized IR-WRI.
We remind that all $p$ sources are inverted at each iteration in the deterministic case, while in the randomized IR-WRI, only $q$ sketched sources are inverted at each iteration. The randomized algorithm reduces to the deterministic algorithm if the sketching matrix is identity, $\bold{\Sketch=I}$. We consider the deterministic inversion result as a reference model to assess the results obtained by the randomized inversion.

\begin{figure}[!t]
\begin{center}
\includegraphics[scale=0.9,trim={0.7cm 0 0 0},clip]{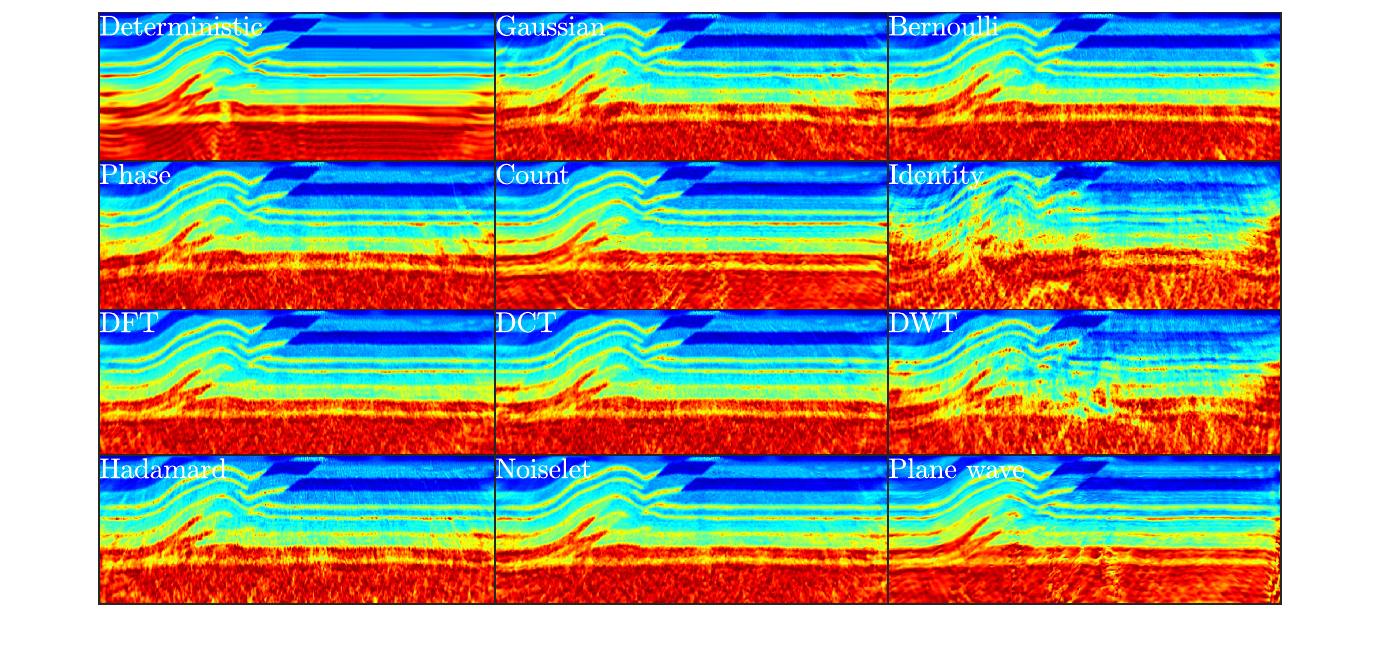}
\end{center}
\caption{Overthrust model. Inversion results from different methods with $p=134$ and $q=2$.} \label{over_q2}
\end{figure}
\begin{figure}[!t]
\begin{center}
\includegraphics[scale=0.9,trim={0.7cm 0 0 0},clip]{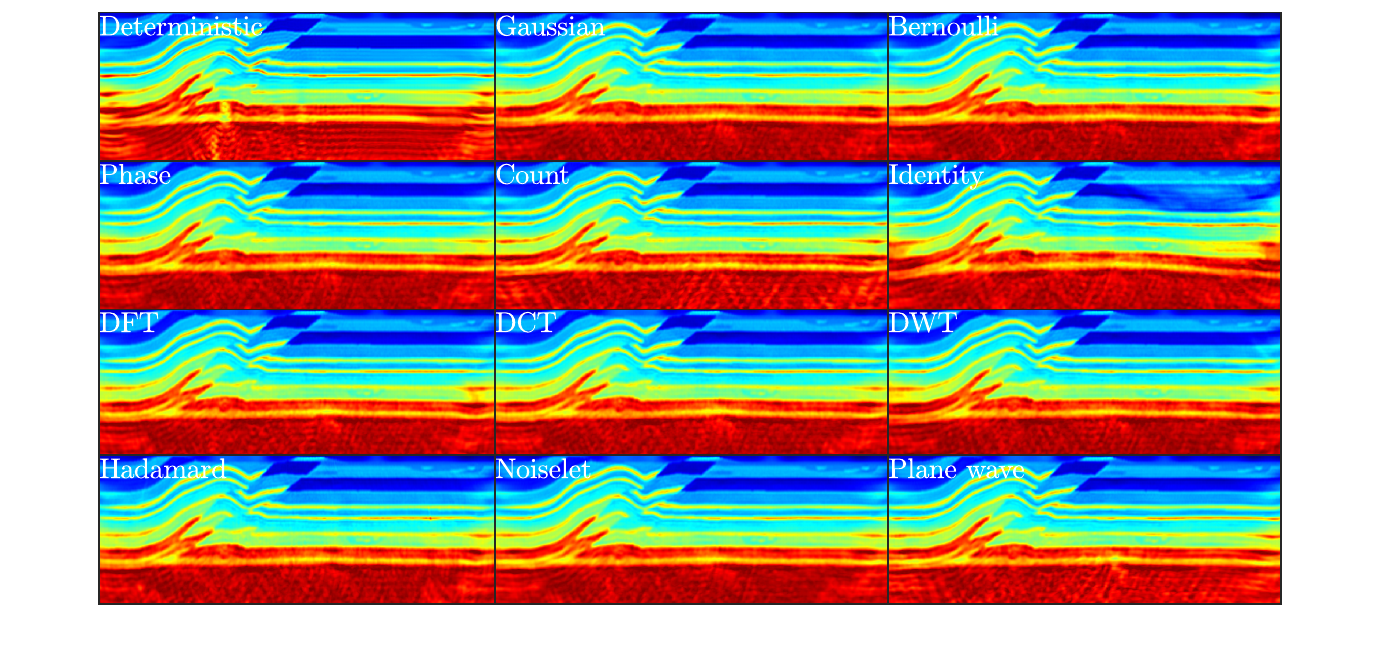}
\end{center}
\caption{Overthrust model. Inversion results from different methods with $p=134$ and $q=10$.} \label{over_q10}
\end{figure}
\begin{figure}[!t]
\begin{center}
\includegraphics[scale=0.7]{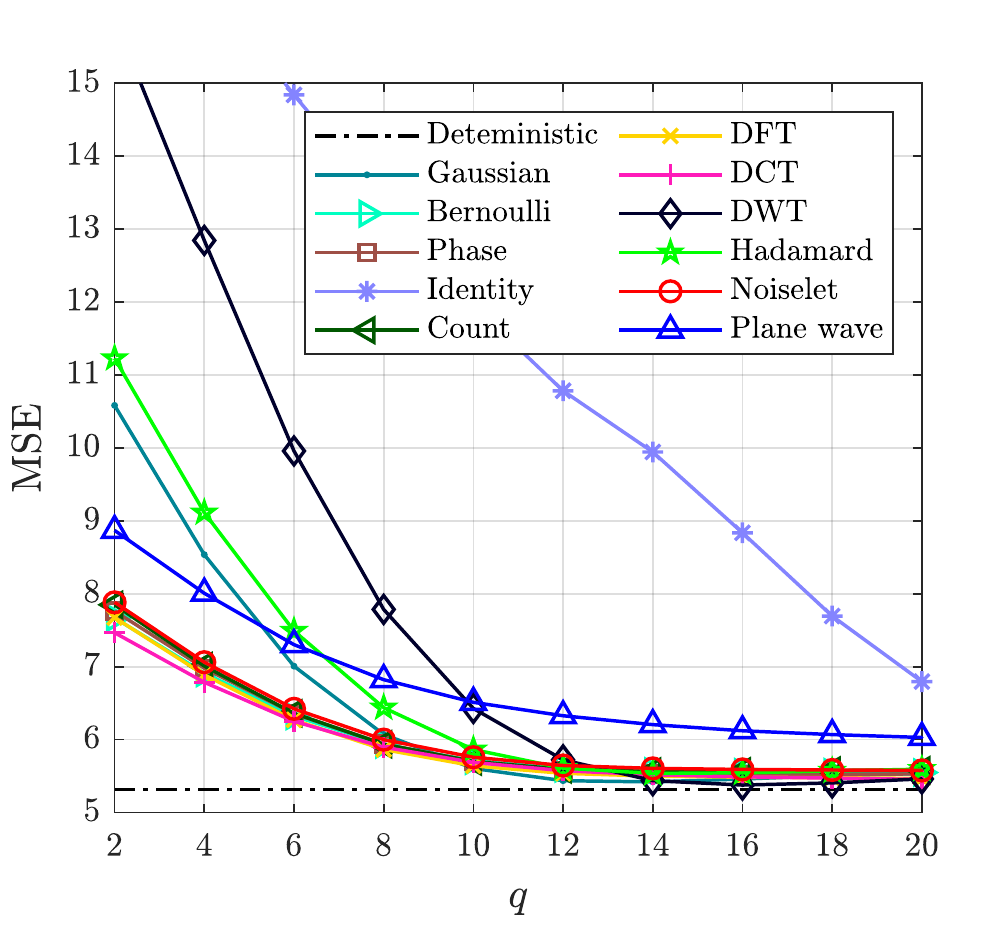}
\end{center}
\caption{The MSE curve versus $q$ for different methods associated to the Overthrust model.} \label{over_mse}
\end{figure}
Figs. \ref{over_q2} and \ref{over_q10} show the inversion result for $q=2$ and $q=10$, respectively, where the deterministic result (with $p=134$) is shown for comparison.  
We see that even with $q=2$, the randomized inversion performs satisfactorily with most of the sketching matrices.
Among them, DCT performed best, and the Identity and DWT performed worst (of all). 
Fig. \ref{over_mse} shows the evolution of the MSE (averaged over 10 realizations) of different methods for a range of $q$ values between 2 and 20. We observe that for large $q$ all error curves approach the deterministic error (shown by horizontal dashed line). Furthermore, sketching with DCT, DFT, Noiselet, Count, Gaussian, and Bernoulli performed almost similarly, and their quality is nearly equal to that of deterministic inversion for $q>12$. 

\begin{figure}[!t]
\begin{center}
\includegraphics[scale=0.9,trim={0.7cm 0 0 0},clip]{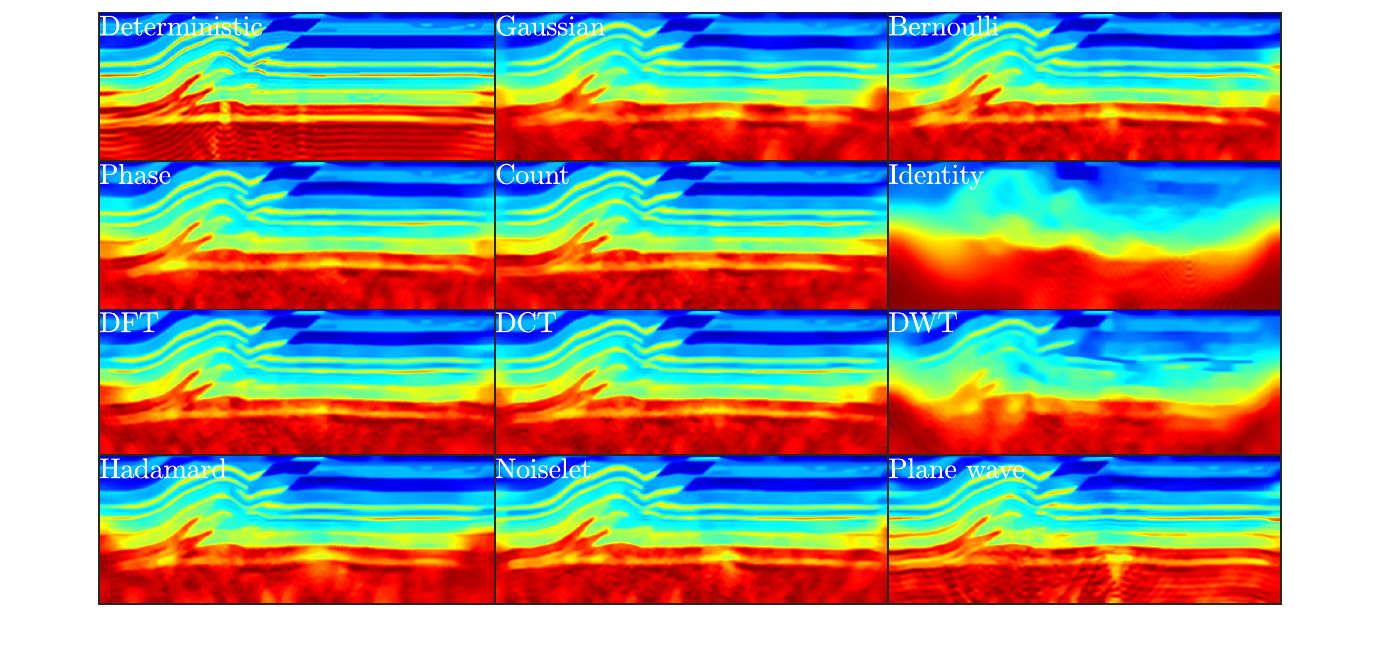}
\end{center}
\caption{Same as Fig. \ref{over_q2} but (BM3D) regularization applied.} \label{over_q2_reg}
\end{figure}

Comparing the results obtained for $q=2$ and $q=10$ (Figs. \ref{over_q2} and \ref{over_q10}) shows that the former results suffer from cross-talk noises which mainly takes the form of random noise in the final model. In order to suppress such noise, we applied the adaptive Block-matching and 3D filtering (BM3D) regularization \cite{Danielyan_2011_BFV, Aghamiry_2020_FWI} for the case $q=2$. The regularized inversion results are shown in Fig. \ref{over_q2_reg}. We observe that the cross-talk noise has been significantly attenuated, showing that we can use a very low number of $q$ while maintaining the quality of the reconstruction with appropriate regularization. 

%
%
%
\begin{figure}[!t]
\begin{center}
\includegraphics[scale=0.9,trim={0.7cm 0 0 0},clip]{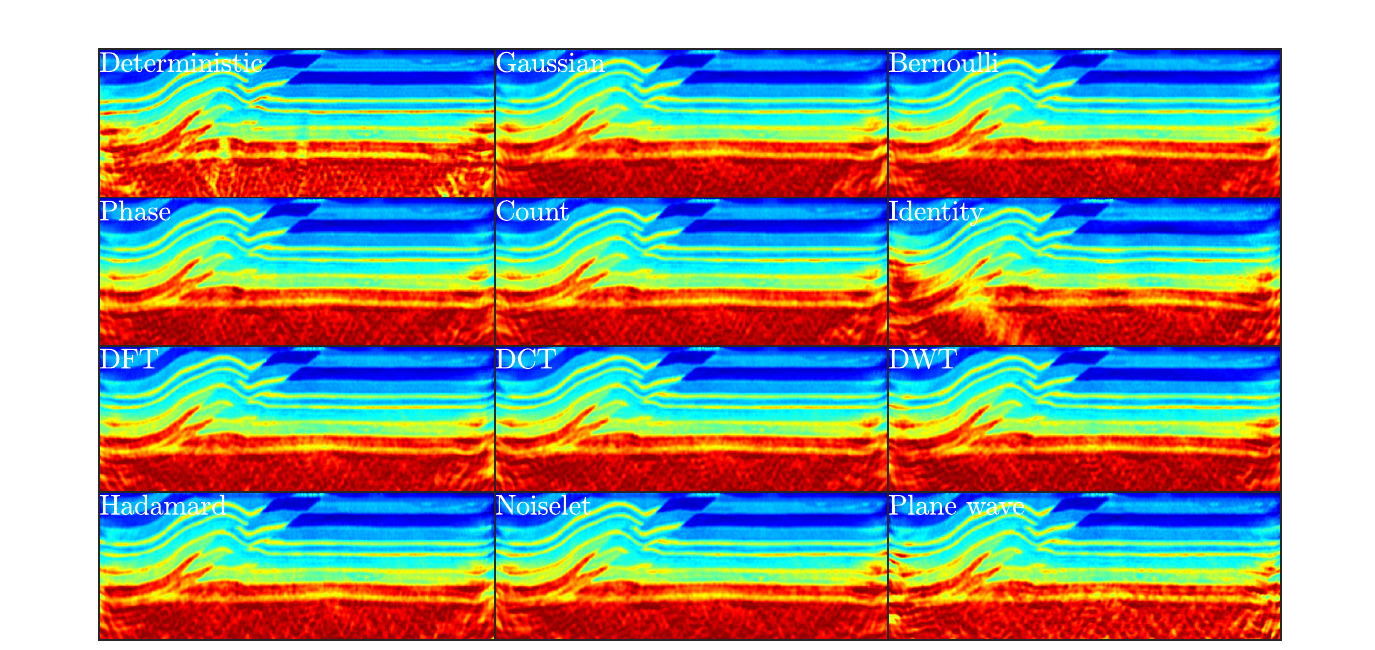}
\end{center}
\caption{Same as Fig. \ref{over_q10} but with noisy data.} \label{over_q10_noise}
\end{figure}

We continued by examining the effect of random noise in the data on the randomized sketching algorithm. For this, the data are contaminated by random noises of signal-to-noise ratio (SNR) 10 dB and inverted with different methods while we set $q=10$. The inversion results are shown in Fig. \ref{over_q10_noise}. We see that the randomized inversion stably generated results with the same quality as those of the deterministic algorithm. In the next subsection, we test the algorithm with the challenging 2004 BP salt model.

\subsection{2004 BP example}
Here we assess the performance of IR-WRI with randomized source sketching against the challenging 2004 BP salt model (Fig. \ref{velmodels}, middle left). The model includes 100 $\times$ 369 grid points with a grid interval of 30~m in both spatial directions. The surface acquisition is conducted with equally spaced $p=62$ sources and $m=185$ receivers. A smooth version of the true velocity model is used as the initial model (Fig. \ref{velmodels}, middle right), and the inversion is performed over two frequency paths, [3-8]~Hz and [3-12]~Hz with a frequency interval of 0.5~Hz in which the final estimated model of the first path is used as an initial model for the second path. Moreover, the maximum number of iterations per frequency is set to 10, except for the first two frequencies for which we perform 25 and 15 iterations, respectively. We perform the inversion with both deterministic and randomized IR-WRI.

\begin{figure}[!t]
\begin{center}
\includegraphics[scale=0.9,trim={0.7cm 0 0 0},clip]{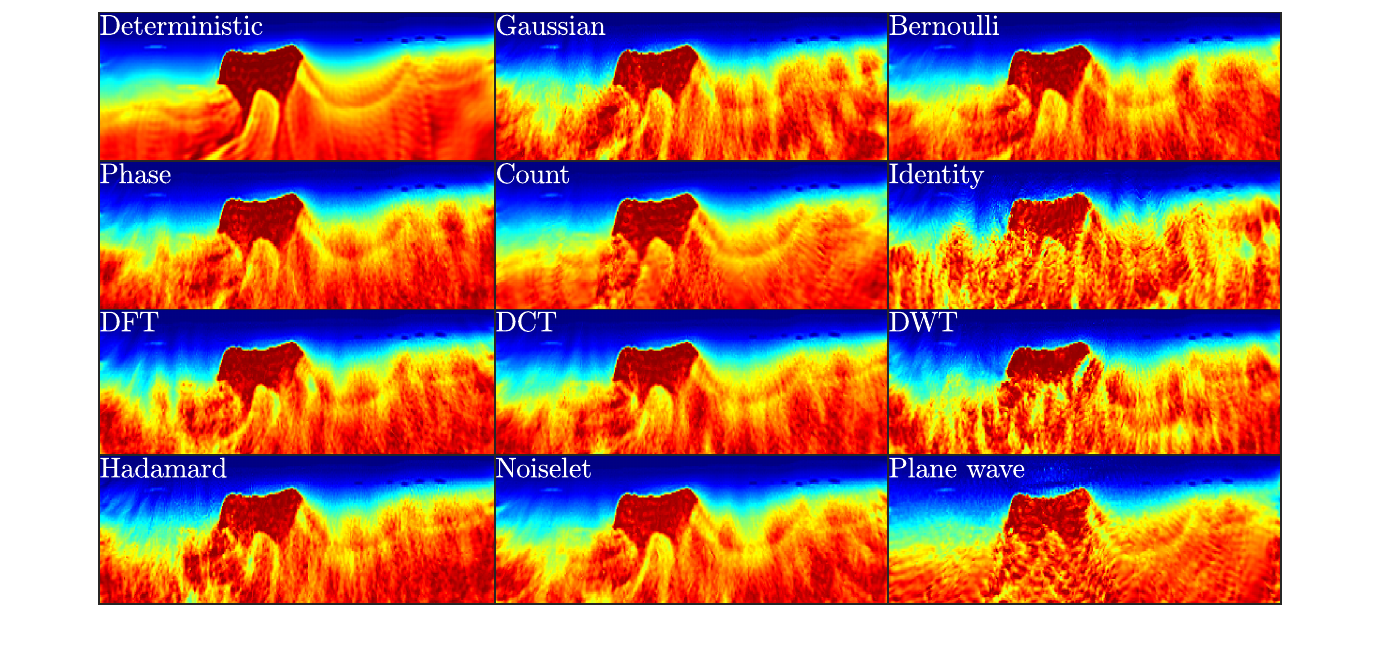}
\end{center}
\caption{Inversion results for the 2004 BP model for different methods with $p=62$ and $q=2$.} \label{bp2004_q2}
\end{figure}
\begin{figure}[!t]
\begin{center}
\includegraphics[scale=0.9,trim={0.5cm 0 0 0},clip]{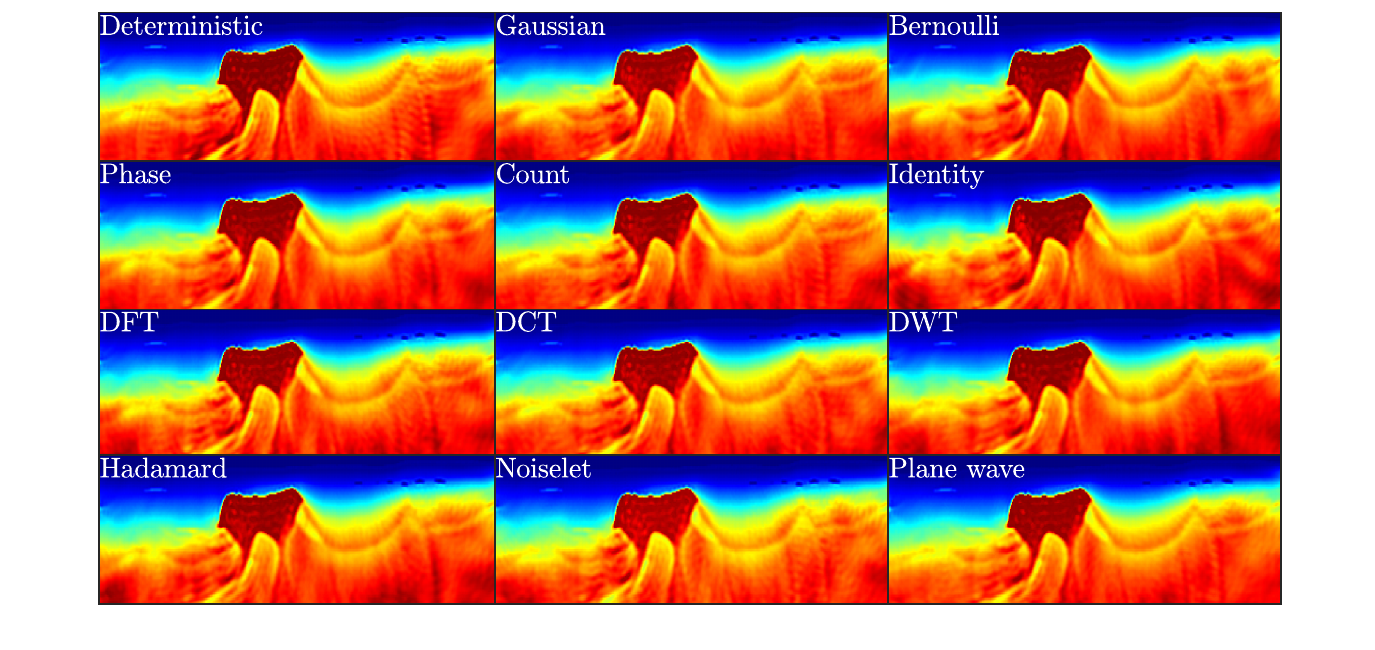}
\end{center}
\caption{Inversion results for the 2004 BP model for different methods with $p=62$ and $q=10$.} \label{bp2004_q10}
\end{figure}
\begin{figure}[!t]
\begin{center}
\includegraphics[scale=0.6]{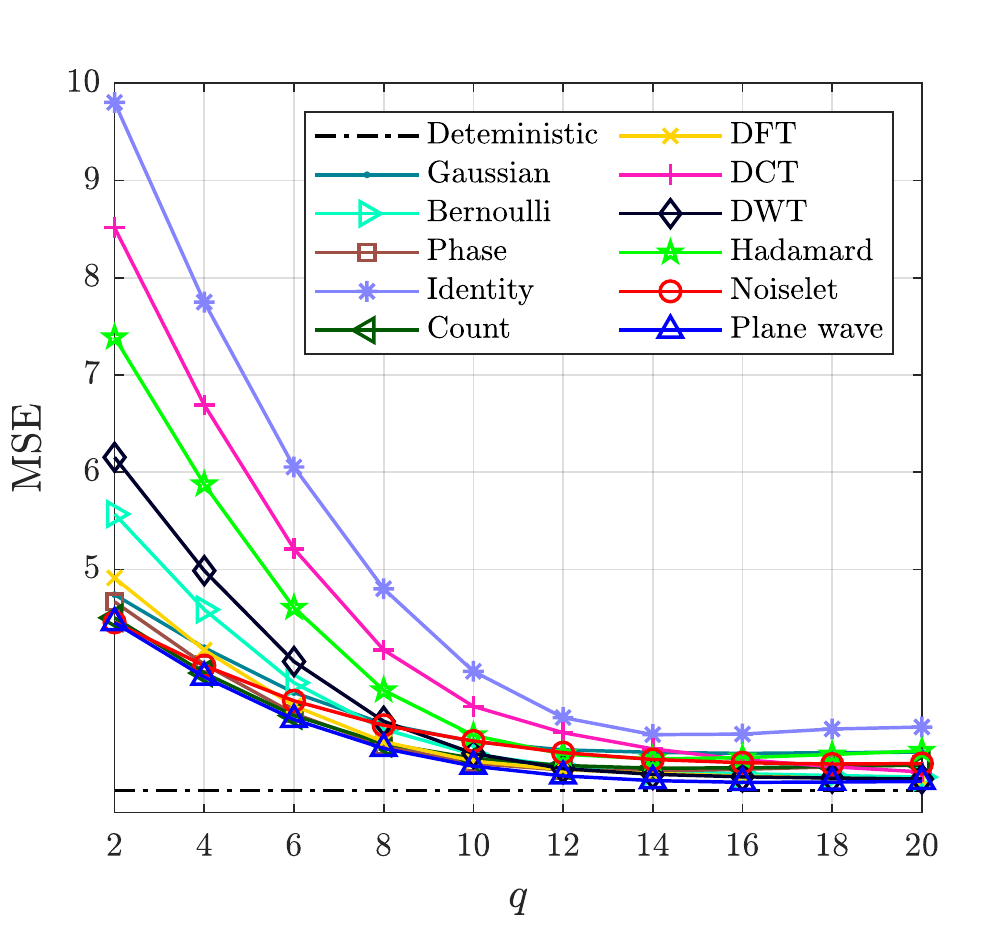}
\end{center}
\caption{The MSE curve versus $q$ for different methods associated to the 2004 BP model.} \label{bp2004_mse}
\end{figure}
\begin{figure}[!t]
\begin{center}
\includegraphics[scale=0.9,trim={0.7cm 0 0 0},clip]{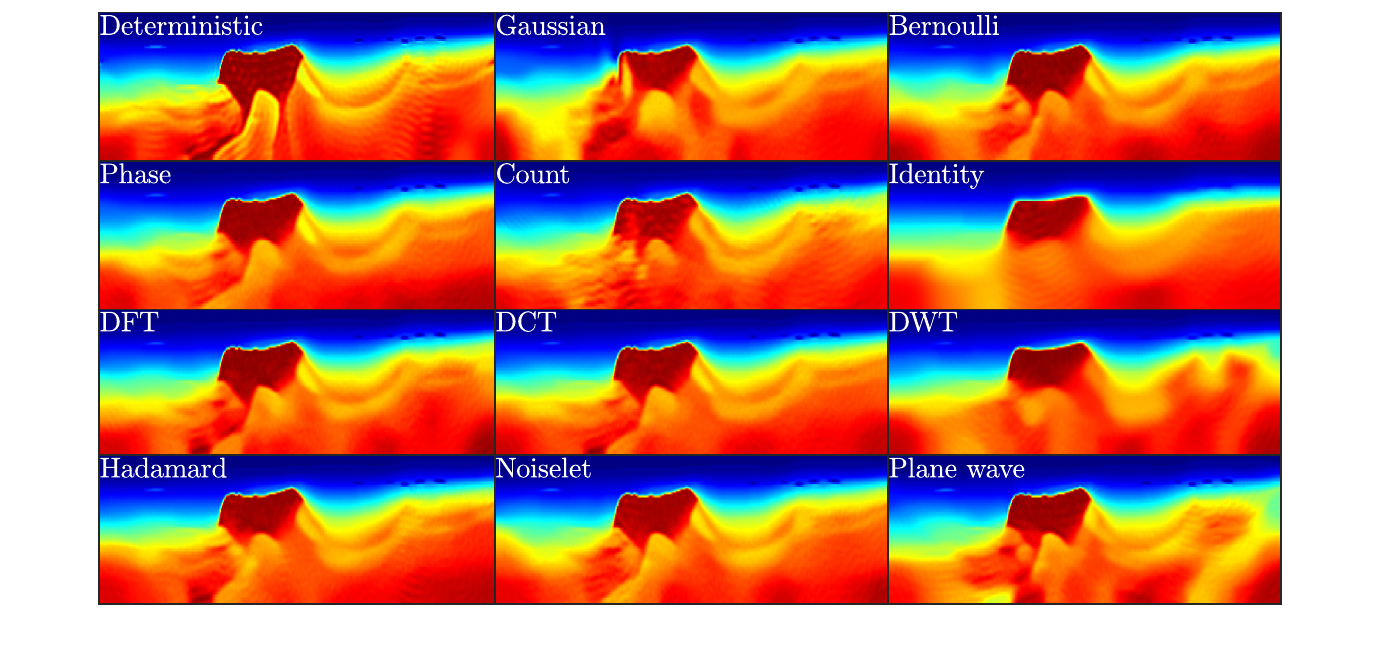}
\end{center}
\caption{Same as Fig. \ref{bp2004_q2} but with (BM3D) regularization applied.} \label{bp2004_q2_reg}
\end{figure}

%
%
%
%
%
%
Figs. \ref{bp2004_q2} and \ref{bp2004_q10} show the randomized inversion results for $q=2$ and $q=10$, respectively, while the deterministic result (with $p=62$) is shown for comparison.  
Again, we see that the randomized inversion with $q=2$ was able to recover the main structures of the model for all of the sketching matrices while all of results obtained with $q=10$ have the same quality as those obtained with the deterministic inversion. 
Fig. \ref{bp2004_mse} shows the evolution of the MSE (averaged over ten realizations) of different methods for a range of $q$ values between 2 and 20. Again, the quality of all methods (except the Identity) is nearly equal to that of the deterministic inversion for $q>12$. 
We also apply the regularization for the case $q=2$ (Fig. \ref{bp2004_q2_reg}). Comparing this figure with Fig. \ref{bp2004_q2} clearly shows the positive impact of regularization in reducing the cross-talk noise.
In the next subsection, we test the algorithm with the 2007 BP salt model.
%
%
%
%

\subsection{2007 BP example}
\begin{figure}[!t]
\begin{center}
\includegraphics[scale=0.9,trim={0.7cm 0 0 0},clip]{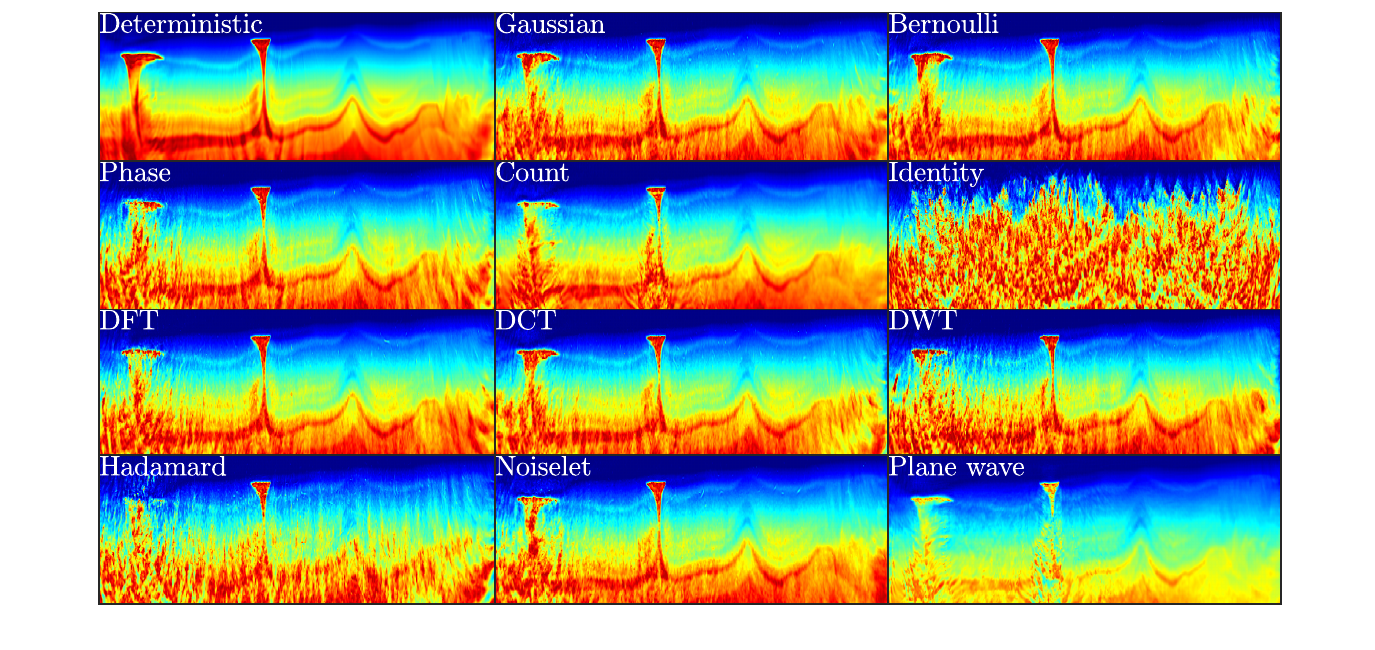}
\end{center}
\caption{Inversion results for the 2007 BP model for different methods with $p=263$ and $q=2$.} \label{bp2007_q2}
\end{figure}

\begin{figure}[!t]
\begin{center}
\includegraphics[scale=0.9,trim={0.7cm 0 0 0},clip]{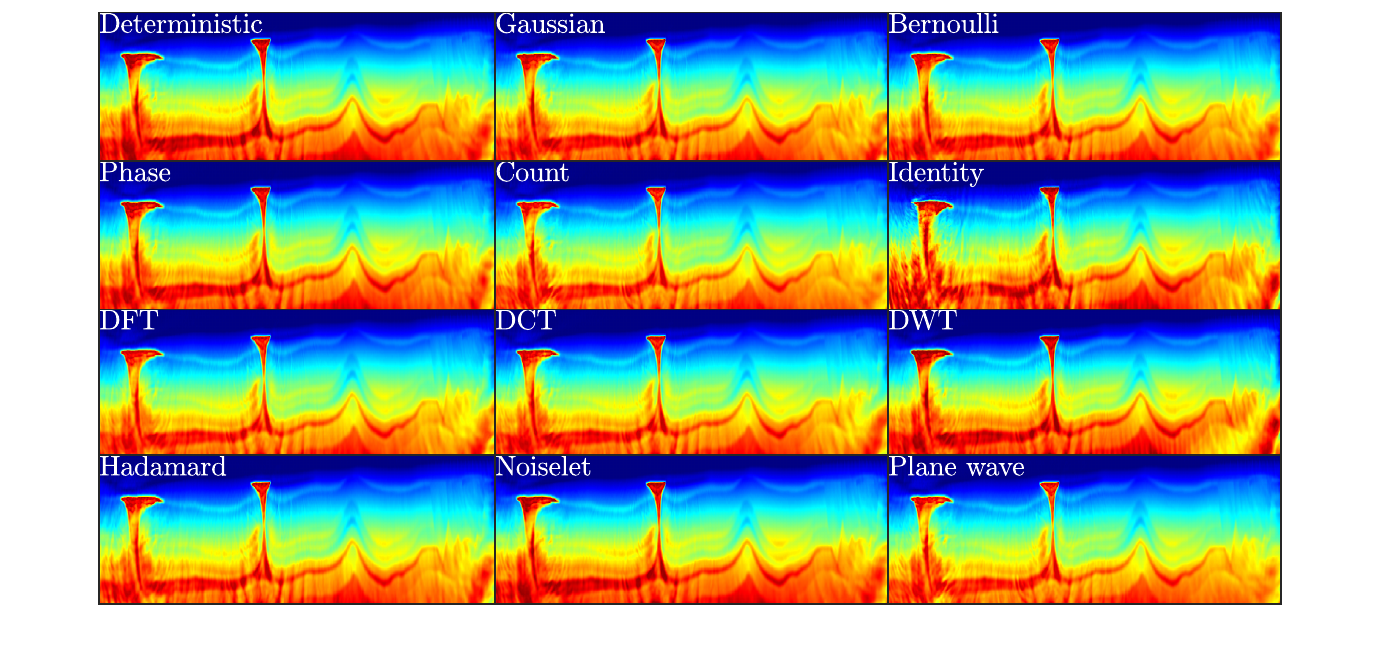}
\end{center}
\caption{Inversion results for the 2007 BP model for different methods with $p=263$ and $q=10$.} \label{bp2007_q10}
\end{figure}

\begin{figure}[!h]
\begin{center}
\includegraphics[scale=0.7]{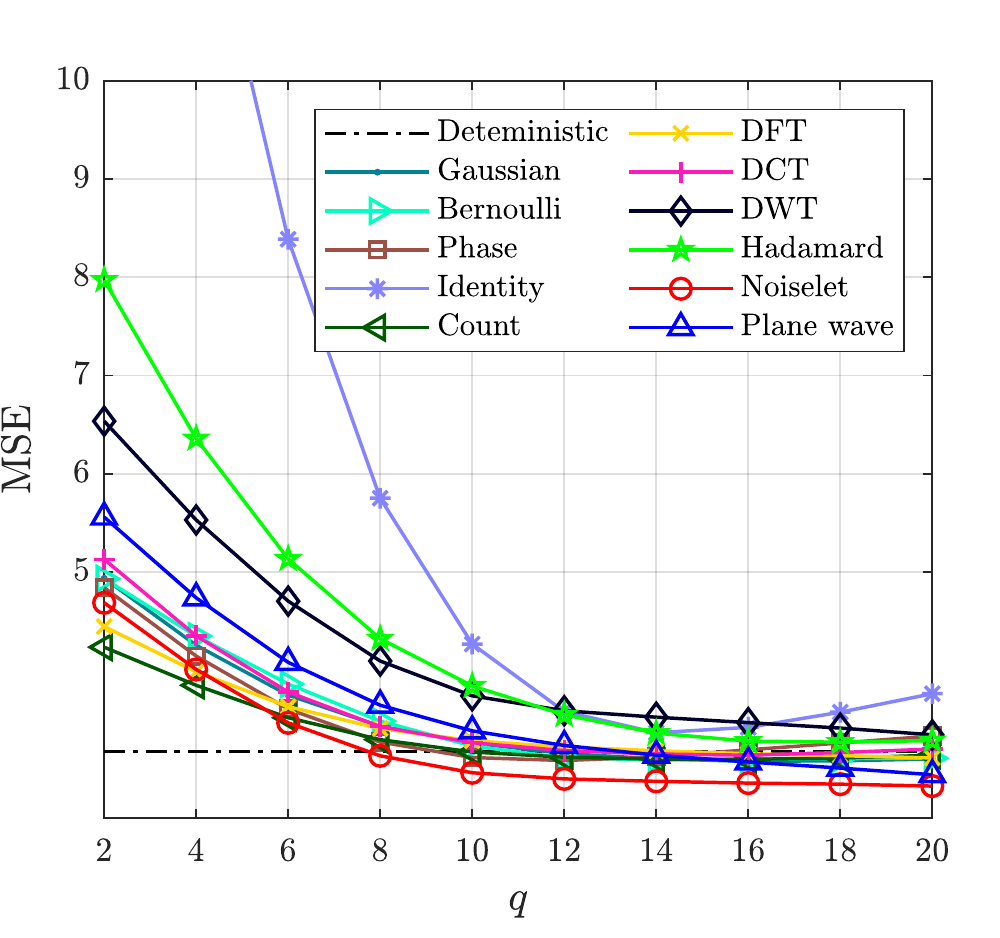}
\end{center}
\caption{The MSE curve versus $q$ for different methods associated to the 2007 BP model.} \label{bp2007_mse}
\end{figure}

Finally, we assess the performance of IR-WRI with randomized source sketching against the large 2007 BP model (Fig. \ref{velmodels}, bottom left). The model includes 151 $\times$ 1050 grid points with a grid interval of 75~m in both spatial directions. The surface acquisition is conducted with equally spaced $p=263$ sources and $m=1050$ receivers. A smooth version of the true velocity model is used as the initial model (Fig. \ref{velmodels}, bottom right), and the inversion is performed over three frequency paths, [1-2]~Hz, [1-3.5] Hz and [1-4]~Hz with a frequency interval of 0.5~Hz in which the final estimated model of each path is used as an initial model for the next path. Moreover, the maximum number of iterations per frequency is set to 10, except for the first frequency for which we perform 20 iterations. We perform the inversion by both deterministic and randomized IR-WRI.
Figs. \ref{bp2007_q2} and \ref{bp2007_q10} show the inversion results for $q=2$ and $q=10$, respectively while the deterministic inversion result (with $p=263$) is shown for comparison.  
This model is more challenging to reconstruct. Generally, the quality of the reconstructions  for $q=2$ is poorer compared to the previous benchmarks. 
The result obtained with the randomized Identity sketching matrix is meaningless. The other methods recover the top of the central salt body, but the left salt is poorly recovered. For $q=10$ (Fig. \ref{bp2007_q10}), however, all methods performed satisfactorily. 
We can see from the MSE (averaged over ten realizations) curves in Fig. \ref{bp2007_mse} that except Identity, Hadamard, and DWT, the error of other methods approaches that of the deterministic for $q>10$. We also see that the randomized Noiselet performs even better than the original deterministic inversion for $q>8$.
Randomized plane wave sketching shows similar behavior for $q>16$.

\begin{figure}[!t]
\begin{center}
\includegraphics[scale=0.9,trim={0.7cm 0 0 0},clip]{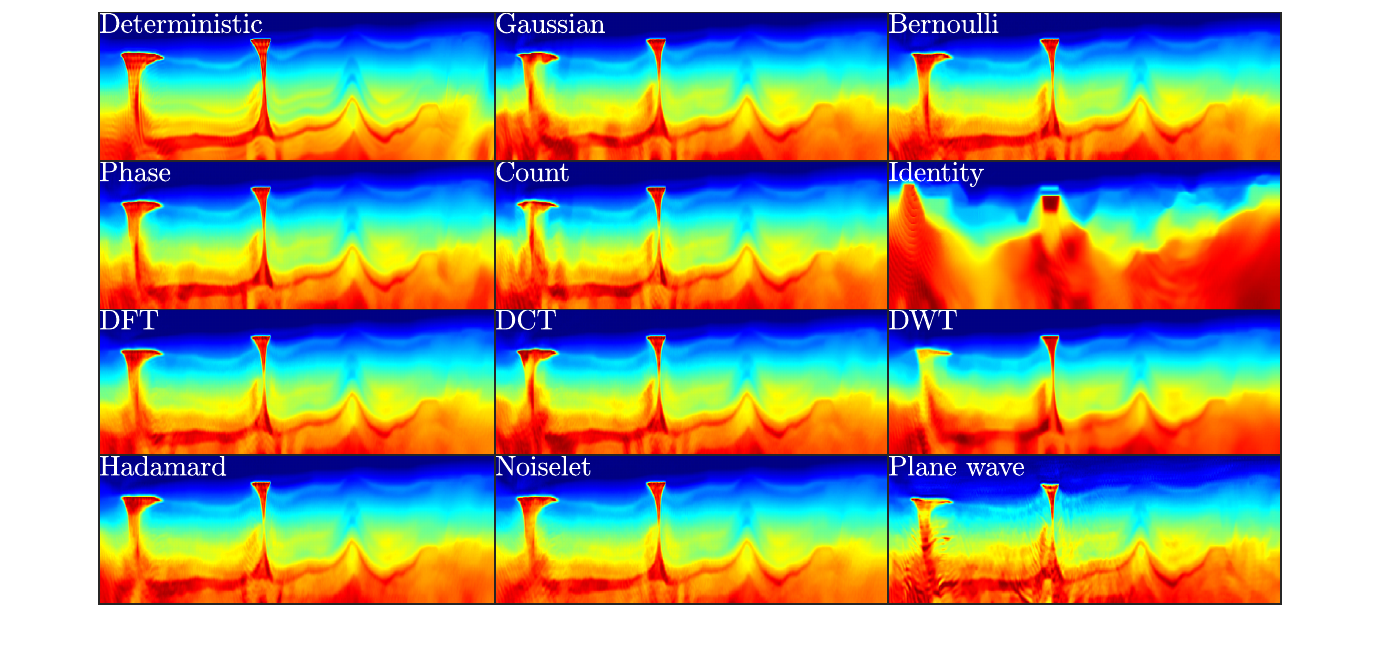}
\end{center}
\caption{Same as Fig. \ref{bp2007_q2} but with (BM3D) regularization.} \label{bp2007_q2_reg}
\end{figure}
\begin{figure}[!t]
\begin{center}
\includegraphics[scale=0.9,trim={0.7cm 0 0 0},clip]{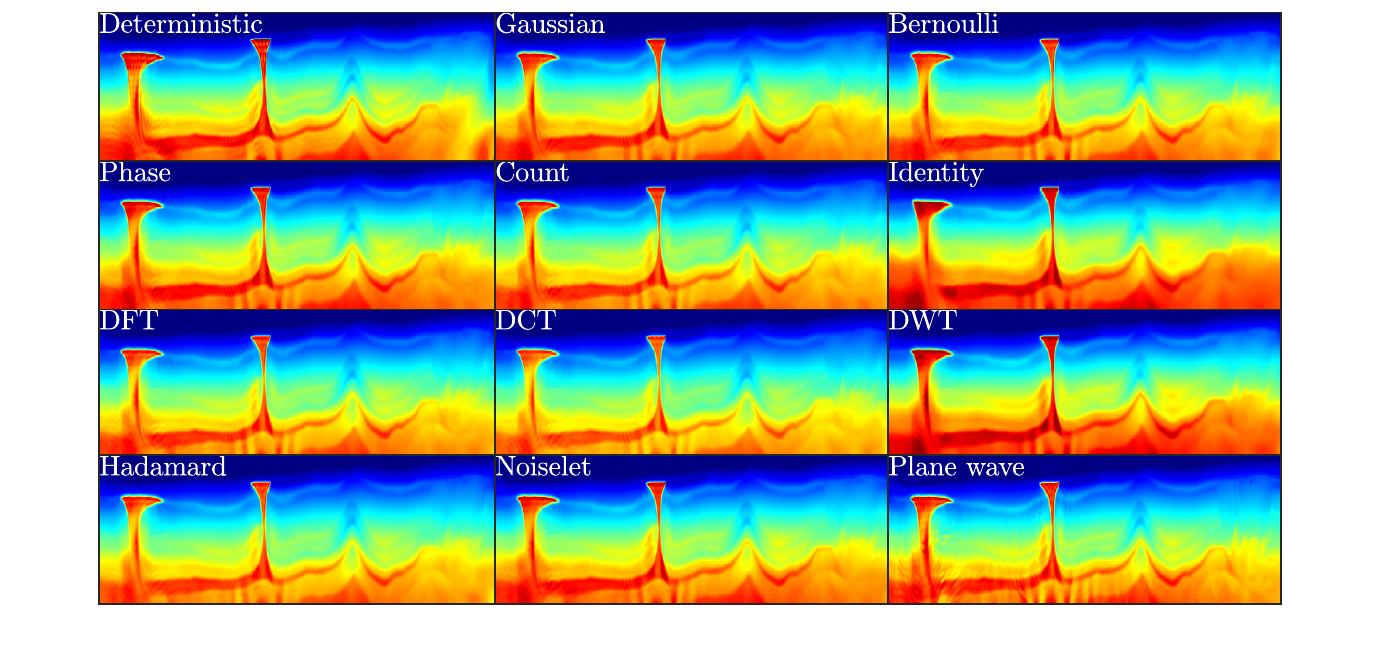}
\end{center}
\caption{Same as Fig. \ref{bp2007_q10} but with (BM3D) regularization.} \label{bp2007_q10_reg}
\end{figure}
\begin{figure}[!h]
\begin{center}
\includegraphics[scale=0.7]{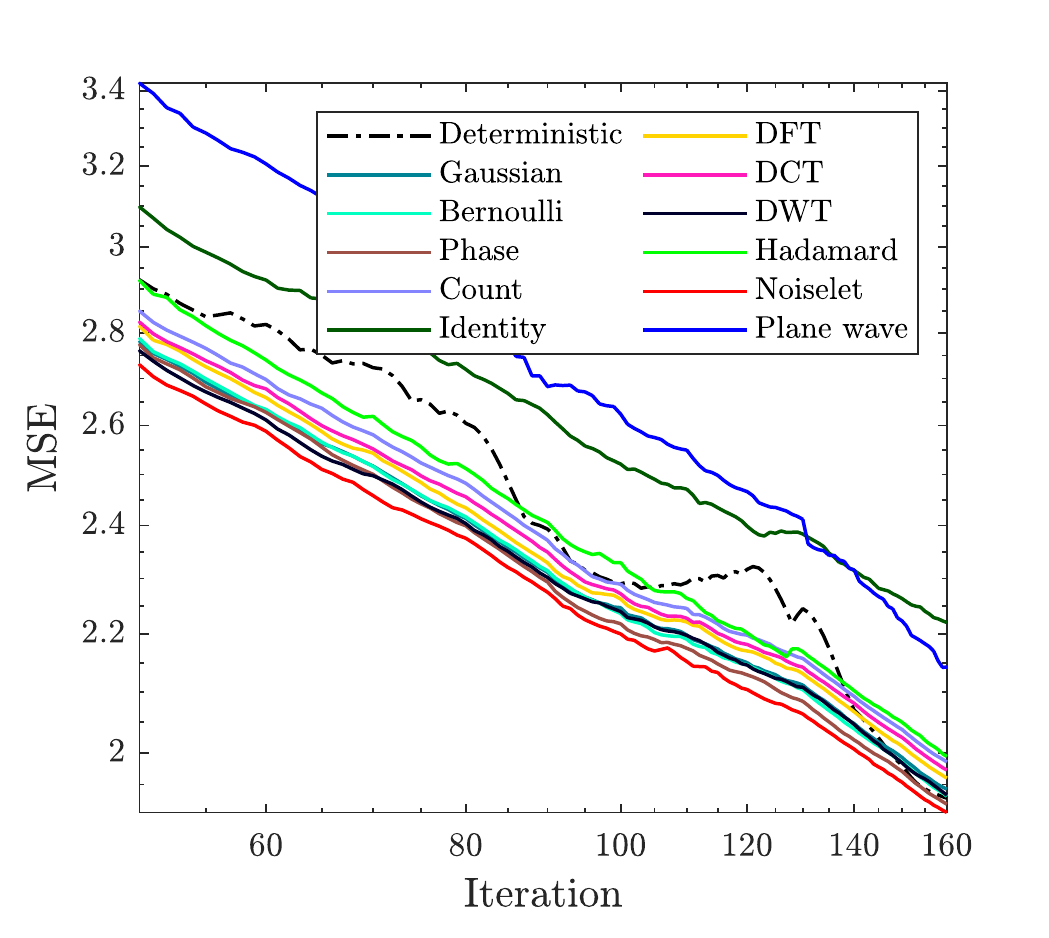}
\end{center}
\caption{The evolution of MSE curve versus IR-WRI iteration for different methods in recovering the inversion results in Fig. \ref{bp2007_q10_reg}.} \label{bp2007_mse_reg}
\end{figure}

In order to examine the effect of regularization for improving the quality of reconstruction for the 2007 BP model, we implemented all the methods for $q=2$ with BM3D regularization (Fig. \ref{bp2007_q2_reg}). For all of the sketching methods, regularization greatly improve the quality (compare Figs. \ref{bp2007_q2} and \ref{bp2007_q2_reg}).  

For a more comprehensive assessment of the regularization, we also apply it with the randomized algorithms for $q=10$. The inversion results after 160 iterations are shown in Fig. \ref{bp2007_q10_reg}. Besides, the associated MSE curves versus iteration are shown in \ref{bp2007_mse_reg}. Interestingly, most of the randomized algorithms with regularization outperformed the original deterministic algorithm with regularization.  This shows that even with less computational burden, the randomized algorithm can explore areas in model space that are not accessible to the deterministic counterpart.

\subsection{Computational Efficiency}
The computational gain of randomized source sketching can be analyzed by the total number of PDEs solves. The total number of PDE solves for each method is equal to the number of inversion iterations multiplied by the number of (right-hand sides) source terms in the data assimilated system. For the deterministic and randomized algorithms, this value equals $N \times p$ and $N \times q$, respectively, wherein $N$ denotes the total number of iterations performed.
Assuming an equal number of iterations, the speedup achieved by the sketching is given by
\begin{equation}
S  = \left(1-\frac{q}{p}\right)\times 100 ~\%/
\end{equation}
The speedup gained for each example presented above is reported in Table \ref{Table:i} and is one to two order of magnitudes.

\begin{table}[ht]
\centering
\caption{Total number of PDE solved and speed up (in percent) for the deterministic IR-WRI and its randomized source sketching variant. } \label{Table:i}
\begin{tabular}{|c|c|c|cccc|} 
\toprule
\multicolumn{1}{c}{\textbf{Model}} & \multicolumn{2}{c}{\textbf{Deterministic}}   & \multicolumn{4}{c}{\begin{tabular}[c]{@{}c@{}}\textbf{Randomized Sketching}\end{tabular}}  \\ 
\hline
                                   & \multicolumn{1}{c}{} &                      &  & $q$=2   &  & $q$=10                                                                                           \\ 
\cline{2-7}
~\textbf{Ovethrust}                & PDE Solved           & 41540                &  & 620   &  & 3100                                                                                           \\ 
\cline{2-7}
                                   & Speed up (\%)        & -                    &  & 98.50 &  & 92.53                                                                                          \\ 
\hline
\multicolumn{1}{c}{}               & \multicolumn{1}{c}{} & \multicolumn{1}{c}{} &  &       &  & \multicolumn{1}{c}{}                                                                           \\ 
\hline
\textbf{2004 BP}                    & PDE Solved           & 21080                &  & 680   &  & 3400                                                                                           \\ 
\cline{2-7}
                                   & Speed up (\%)        & -                    &  & 96.77 &  & 83.87                                                                                          \\ 
\hline
\multicolumn{1}{c}{}               & \multicolumn{1}{c}{} & \multicolumn{1}{c}{} &  &       &  & \multicolumn{1}{c}{}                                                                           \\ 
\hline
\textbf{2007 BP}                    & PDE Solved           & 42080                &  & 320   &  & 1600                                                                                           \\ 
\cline{2-7}
                                   & Speed up (\%)        & -                    &  & 99.23 &  & 96.20                                                                                          \\
\bottomrule
\end{tabular}
\end{table}
\section{Conclusions}
We improve the computational efficiency of an extended-space formulation of FWI based on wavefield reconstruction (iteratively-refined wavefield reconstruction inversion, IR-WRI) using randomized source sketching or encoding. The source dimension of the problem is projected to a lower-dimensional space by using a random sketching matrix, which is regenerated at each iteration. This allowed reducing the number of working sources significantly compared to the original problem, and hence the number of PDE solutions. We showed that the proposed algorithm is quite general and includes the existing source encoding methods as special cases. Accordingly, we proposed a randomized variant of the traditional deterministic plane-wave encoding.  
We assess the performance of the proposed method with three 2D benchmark models. We conclude that the proposed randomized algorithm reduces the computational cost of the wavefield inversion by more than one order of magnitude for noise-free and noisy data. Higher speedup may be obtained for 3D acquisitions. 
We showed that sparsity-promoting regularization can reduce the cross-talk noise which arise when the source dimension is reduced significantly. 
Furthermore, regularization makes the randomized algorithm able to explore areas of the model space that are not accessible to the deterministic counterpart. 

\section{Acknowledgments}  
This study was partially funded by the WIND consortium (\textit{https://www.geoazur.fr/WIND}), sponsored by Chevron, Shell and Total. The authors are grateful to the OPAL infrastructure from 
Observatoire de la Côte d'Azur (CRIMSON) for providing resources and support. This work was granted access to the HPC resources of IDRIS under the allocation A0050410596 made by GENCI.
%
\appendices
\section{Plane wave encoding} \label{PW_sec}
We review the plane wave theory (linear Radon  transform) and construction of the associated projection matrix. It has been used for encoding seismic sources in FWI, and WRI problems in a deterministic setting  \cite{Vigh_2008_PPF,zhang2005delayed,tao2013frequency,kim2020seismic}.  
For a time domain seismic data $d(t, x)$ with spatial variable $x$ (e.g. the source coordinate and in this case $d(t, x)$ will be a common receiver gather) and time $t$ the linear Radon transform of $d$ can be defined by the following integral \cite{Beylkin_1987_DRT}
\begin{equation} \label{RT_t}
\hat{d}(\tau, p)=\int d(t=\tau+\bar{p}x, x)dx
\end{equation}
where $\tau$ is intercept time and $\bar{p}$ is a ray parameter. The integration is computed over the range of $x$. Due to the time invariance nature of this integral equation
it can be computed for each frequency slice $\omega$ separately. After a proper discretization of the equation we get for a single frequency slice of the Radon coefficient
\begin{equation} \label{RT_t}
\hat{d}(\omega,\bar{p})=\sum_x d(\omega,x)e^{\iota\omega \bar{p} x}.
\end{equation}
Accordingly, the augmented Lagrangian \eqref{AL_trans} in the plane-wave domain is obtained by defining the sketch matrix $\bold{\Sketch}$ as
\begin{equation} \label{PW}
\bold{\Sketch}(i,j)=\left\{\begin{array}{ll}
e^{\iota \omega \bold{\bar{p}}(j)(\bold{x}_s(i)-\bold{x}_{\min})} &if~ \bold{\bar{p}}(j) \geq 0, \\
e^{\iota \omega \bold{\bar{p}}(j)(\bold{x}_s(i)-\bold{x}_{\max})} &if~ \bold{\bar{p}}(j)<0, 
\end{array}\right.
\end{equation}
\cite{tao2013frequency}, where $\bold{\bar{p}}$ denotes the ray parameter vector and $\bold{x}_{\min}$/$\bold{x}_{\max}$ denote the minimum/maximum offset. 
Using the encoding matrix \eqref{PW}, the $\Ns$ original sources are combined linearly to be replaced by $\Np$ plane-wave sources, where $\Np$ is the number of ray parameters. 
The success of the plane-wave encoding relies on the range and number of ray parameters which can be determined by the following condition \cite{zhang2005delayed}:
\begin{equation}
\Np \geq x_{\max} f_{\max } (\bar{p}_{\max}-\bar{p}_{\min}),
\end{equation}
where $\Np$ is the minimum required number of ray parameters, $x_{\max}, f_{\max}$ denote the maximum offset and frequency, respectively, and  $\bar{p}_{\max}/\bar{p}_{\min}$ denote the maximum/minimum ray parameter.

\bibliographystyle{IEEEtran}
\newcommand{\SortNoop}[1]{}

\end{document}